\documentclass[11pt, reqno]{amsart}
\usepackage{amsmath, amsthm, a4, latexsym, amssymb}

\def\@rmrk#1#2{\refstepcounter
    {#1}\@ifnextchar[{\@yrmrk{#1}{#2}}{\@xrmrk{#1}{#2}}}

\makeatletter\@addtoreset{equation}{section}\makeatother

 \newfont{\bfit}{cmbxti10 scaled 2000}
 \newfont{\biggi}{cmr12 scaled 2000}

 \newcommand{\eps}{\varepsilon}

 \newcommand{\R}{\mathbb{R}}
 \newcommand{\Z}{\mathbb{Z}}
 \newcommand{\N}{\mathbb{N}}

 \newcommand{\prob}{\mathbb{P}}

 \newcommand{\me}{\mathbb{E}}
 
 \renewcommand{\P}{\mathbb{P}}
 
 \newcommand{\one}{\1}
 \newcommand{\tmu}{\widetilde{\mu}}

 \newcommand{\tome}{\tilde{\omega}}

 \newcommand{\skric}{{\mathcal C}}

 \newcommand{\skrie}{{\mathcal E}}
 
 \newcommand{\skrig}{{\mathcal G}}

 \newcommand{\skrim}{{\mathcal M}}
 
 \newcommand{\skrit}{{\mathcal T}}

 \newcommand{\skrix}{{\mathcal X}}
 \newcommand{\skriy}{{\mathcal Y}}
 
 \newcommand{\heap}[2]{\genfrac{}{}{0pt}{}{#1}{#2}}
 \newcommand{\sfrac}[2]{\mbox{$\frac{#1}{#2}$}}

\def\1{{\mathchoice {1\mskip-4mu\mathrm l}      
{1\mskip-4mu\mathrm l}
{1\mskip-4.5mu\mathrm l} {1\mskip-5mu\mathrm l}}}

\newcommand{\eq}{\begin{equation}}
\newcommand{\en}{\end{equation}}

\newenvironment{Proof}
{\vskip0.1cm\noindent{\bf Proof. }{\hspace*{0.3cm}}}%
{\nopagebreak {\hspace*{\fill}\rule{2mm}{2mm}}\\ }

{\nopagebreak {\hspace*{\fill}\rule{2mm}{2mm}}\\ }

\renewcommand{\subsection}{\secdef \subsct\sbsect}
\newcommand{\subsct}[2][default]{\refstepcounter{subsection}
\vspace{0.15cm}
{\flushleft\bf \arabic{section}.\arabic{subsection}~\bf #1  }
\nopagebreak\nopagebreak}
\newcommand{\sbsect}[1]{\vspace{0.1cm}\noindent
{\bf #1}\vspace{0.1cm}}

\newtheorem{theorem}{Theorem}[section]
\newtheorem{lemma}[theorem]{Lemma}

\newtheoremstyle{thm}{1.5ex}{1.5ex}{\itshape\rmfamily}{}
{\bfseries\rmfamily}{}{2ex}{}

\newtheoremstyle{rem}{1.3ex}{1.3ex}{\rmfamily}{}
{\itshape\rmfamily}{}{1.5ex}{}
\theoremstyle{rem}
\newtheorem{remark}{{\slshape\sffamily Remark}}[]

\refstepcounter{subsection}

\def\thebibliography#1{\section*{References}
  \list%
  {\arabic{enumi}.}
    {\settowidth\labelwidth{[#1]}\leftmargin\labelwidth
    \advance\leftmargin\labelsep
    \parsep0pt\itemsep0pt
    \usecounter{enumi}}
    \def\newblock{\hskip .11em plus .33em minus .07em}
    \sloppy                   
    \sfcode`\.=1000\relax}

\begin{document}
\title[AEP for hierarchical and networked structures]
{\Large Asymptotic Equipartition Properties for simple hierarchical
and networked structures}

\author[Kwabena Doku-Amponsah]{}

\maketitle
\thispagestyle{empty}
\vspace{-0.5cm}

\centerline{\sc{By Kwabena Doku-Amponsah}}
\renewcommand{\thefootnote}{}
\renewcommand{\thefootnote}{1}
\renewcommand{\thefootnote}{}
\footnote{\textit{Address:} Statistics Department, University of
Ghana, Box LG 115, Legon,Ghana.\,
\textit{E-mail:\,kdoku@ug.edu.gh}.}
\renewcommand{\thefootnote}{1}
\centerline{\textit{University of Ghana}}

\begin{quote}{\small }{\bf Abstract.}
We prove \emph{asymptotic equipartition properties} for simple
hierarchical structures (modelled as \emph{multitype Galton-Watson
trees}) and networked structures (modelled as \emph{randomly
coloured random graphs}). For example, for large~$n$, a networked
data structure consisting of $n$ units connected by an average
number of links of order  $n/\log n$ can be coded by about $H\times
n$ bits, where $H$ is an explicitly defined entropy. The main
technique in our proofs are large deviation principles for suitably
 defined empirical measures.
\end{quote}\vspace{0.5cm}

\section{Introduction}
Information is often  structured in a nonlinear way. For example, in
genetics information often has an implicit hierarchical structure,
in computer science data is often organized in the form  of a
network. To transmit or compress data from these sources, one needs
efficient coding schemes and approximate pattern matching
algorithms, and the \emph{Shannon-McMillan-Breiman theorem} or
\emph{asymptotic equipartition property }(AEP) plays a key role in
this regard, for example by providing bounds on the possible
performance of algorithms.

Two major sets of research work on the AEP (and its applications)
within  mathematics and information theory have so far been
considered. The first of these  has focussed on stationary ergodic
processes such as Markov  chains, see Cover and Thomas \cite{CT91}
and the references therein. The second has dealt with stationary
(ergodic) random fields on $\Z^{d},$ as well as amenable group
actions, see, for example Dembo and Kontoyiannis \cite{DK02} and the
reference therein. Whilst typical examples of applications of the
former has concentrated on data from linear source, the latter
includes recent advances such as image and video processing,
geostatistics, and statistical mechanics.

However, numerous types of data we usually come across in
applications (communication studies, demographic studies, biological
population studies and the field of physics) are naturally
structured like networks or trees. For example, the WWW (consisting
of a collection of pages residing on a server with a given name,
together with `hyperlinks' with their direction ignored), data on
the spread of  some disease in a given population and many more, can
be described by networks. Equally, the age structure of a given
population is best modelled by genealogical trees.

In this paper we use the large deviation techniques, as provided in
the recent paper Dembo, M\"{o}rters and Sheffied ~\cite{DMS03}, 
to study the AEP of structured
data consisting of a large number of \emph{units}, chosen from a
finite set, together with a number of \emph{links} connecting the
units.

As an application of our abstract principles, we consider the
following concrete examples from biology.

\begin{itemize} \item {\bf Metabolic network:} This is a
graph of interactions forming a part of the energy generation and
biosynthesis metabolism of the bacterium E.coli. Here, the units
represent \emph{substrates} and \emph{products}, and links represent
\emph{interactions}. See Newman \cite{New00}.

\item{\bf Mutation study:} Consider mutations
in mitochondrial DNA (mtDNA for short) e.g. the mtDNA$^{4977}$
deletion (a mutation which causes a deletion of about one third of
the mitochondrial genome). The  replication of mtDNA can be
described by a tree, where the units are $a$ (normal) and $b$
(mutant) and the links indicate `mother-child' relations. See
Olofsson and Shaw \cite{OS02} and the references therein.

\end{itemize}

The core results of the paper are the \emph{Shannon-McMillan-Breiman
theorems} for two simple probabilistic models: The \emph{multitype
Galton-Watson trees} describing hierarchical data structures, and a
class of \emph{randomly coloured random graphs} describing networked
data structures, see
Theorems~\ref{randomgsmb.sparse}~and~\ref{smb.tree}.

Specifically, we consider for the first model typed trees described
by the following procedure: The root carries a random type chosen
according to the some law on a finite alphabet; given the type of a
vertex, the number and types of the children (ordered from left to
right) are given independently of everything else, by an offspring
law. For the second  we look at random graph models constructed as
follows: Assign vertices colours independently and identically
according to some colour law on a finite set of colours; connect any
pair of vertices independently according to a probability depending
on their colours.This model, with the  simple Erd\H os-R\'enyi graph
with independent colours as a special case, was introduced by Penman
in his thesis~\cite{Pe98}, see Canning and Penman \cite{CP03} for an
exposition.

We also present large deviation principles (LDPs) for empirical
colour measure and empirical pair measure of  \emph{sub}-and
\emph{supercritical} coloured random graphs. Major tool used in the
proofs of these LDPs is (exponential) \emph{change of measure}. We
remark here that some of our results fit well into the framework of
large deviations for mixtures which is utilized in the proofs  of
the LDPs in Doku  and M\"{o}rters \cite{DM06a}.


\subsection{A model for simple hierarchical structures.}\label{SHS}
We review in this subsection, the model for simple hierarchical data
structures,\emph{ multitype Galton-Watson trees}.  To begin, we
collect some notation and concepts from the paper Dembo et al.
\cite{DMS03}. By $\skrit$ we denote the set of all finite rooted
planar trees $T$, by $V=V(T)$ the set of all vertices and by
$E=E(T)$ the set of all edges oriented away from the root, which is
always denoted by $\rho$. We write $|T|$ for the number of vertices
in the tree $T.$ Let $\skrix$ be a finite alphabet and write
$$\displaystyle\skrix^*=\bigcup_{n=0}^\infty \{n\} \times
\skrix^n.$$ We equip $\skrix^*$  with the finest topology with all
subsets as open sets. i.e. the discrete topology. We observe that
the offspring of any vertex $v\in T$ is characterized by an element
of $\skrix^*$ and that there is an element $(0,\emptyset)$ in
$\skrix^*$ symbolizing absence of offspring.

Let $\mu$ be a probability measure (initial distribution) on
$\skrix$ and ${\mathbb{Q}}:\skrix\times\skrix^{*}\rightarrow[0,\,1]$
be an offspring transition kernel. The law $\prob$ of a tree-indexed
process $X$ is defined by the following procedure:
\begin{itemize}
\item The root $\rho$ carries a random type $X(\rho)$ chosen according
to the probability measure $\mu$ on $\skrix.$
\item For every vertex with type $a\in\skrix$ the offspring number
and types are given independently of everything else, by the
offspring law ${\mathbb{Q}}\{\cdot\,|\,a\}$  on $\skrix^{*}.$ We
write
$$
{\mathbb{Q}}\big\{\cdot\,|\,a\}={\mathbb{Q}}\big\{(N,X_1,...,X_N)\in\,\cdot\,|\,a\big\},$$
ie we have a random number $N$ of descendants  with types
$X_1,...,X_N.$
\end{itemize}
We shall consider $X=((X(v),\,C(v)),\,v\in V)$  under the joint law
of tree and offspring. We interpret $X$ as \emph{multitype
Galton-Watson tree} and $X(v)$ as the type of vertex $v.$ For each
typed tree $X$ and each vertex $v$ we denote by $\displaystyle
C(v)=(N(v),X_1(v),\ldots,X_{N(v)}(v))\in \skrix^*,$ the number and
types of the children of $v$, ordered from left to right. We notice
that the children of the root (denoted by $\rho$) are ordered but
the root itself is not. We call an offspring distribution
${\mathbb{Q}}$ bounded if for some $\,N_0<\infty,$ we have
$${\mathbb{Q}}\{N>N_0\,|\,a\}=0, \, \mbox{ for all $\,a\in\skrix\,$.}$$
Denote, for every $c=(n(c),a_1(c),\ldots,a_n(c))\in\skrix^{*}$ and $
a\in\skrix,$ the \emph{multiplicity} of the symbol $a$ in $c$ by
 $$m(a,c)=\sum_{i=1}^{n(c)}1_{\{a_{i}=a\}}.$$

Define the matrix $A$ with index set $\skrix\times\skrix$ and
nonnegative entries by
$$A(a,b)=\sum_{c\in\skrix^{*}}{\mathbb{Q}}\{c\,|\,b\}m(a,c),\mbox{
for $a,b\in\skrix.$}$$

$A(a,b)$ is the expected number of offspring of type $a$ of a vertex
of type $b.$  Let
$A^{*}(a,b)=\sum_{k=1}^{\infty}A^{k}(a,b)\in[0,\infty].$ We say that
the matrix $A$ is irreducible if $A^{*}(a,b)>0,$ for all
$a,b\in\skrix.$

The multitype Galton-Watson tree is called irreducible if the matrix
$A$ is irreducible. It is called critical $($subcritical,
supercritical$)$ if the largest eigenvalue of the matrix $A$ is $1 $
$($ less than $1,$ greater $1$ resp.$).$ Let $\pi$ be the
\emph{eigenvector} corresponding to the largest \emph{eigenvalue} of
the matrix $A$ (normalized to a probability vector).  Then $\pi$ is
\emph{unique}, if the Galton-Watson tree is irreducible.

\subsection{A model for simple networked structures.}\label{SNS}In this subsection,
we review the model for simple networked structures, the
\emph{randomly coloured random graph model}. We begin by fixing the
following notations. Let $\skrix$ be a finite alphabet or colour set
$\skrix.$ Let  $V$  be  a fixed set of $n$ vertices, say
$V=\{1,\ldots,n\}.$ Denote by $\skrig$ the set of all (simple)
graphs and by $\skrig_n$ the set of all (simple) graphs  with vertex
set $V=\{1,\ldots,n\}$ and edge set
$$E\subset\skrie:=\big\{(u,v)\in V\times V \, : \, u<v\big\},$$
where the \emph{formal} ordering of edges is introduced as a means
to simply describe \emph{unordered} edges.

Given a symmetric function $p_n\colon\skrix\times\skrix\rightarrow
[0,1]$ and a probability measure $\mu$ on $\skrix$ we may define the
{\em randomly coloured random graph} or simply \emph{coloured random
graph}~$X$ with $n$ vertices as follows: Assign to each vertex $v\in
V$ colour $X(v)$ independently according to the {\em colour law}
$\mu.$  Given the colours, we connect any two vertices $u,v\in V$,
independently of everything else,  with a {\em connection
probability} $p_n(X(u),X(v))$ otherwise keep them disconnected.  We
always consider $X=\big((X(v)\,:\,v\in V),E\big)$ under the joint
law of graph and colour. We interpret $X$ as coloured random graph
and consider $X(v)$ as the colour of the vertex $v$. Denote by
$\skrig_n(\skrix)$ the set of all coloured graphs with colour set
$\skrix$ and $n$ vertices.

We look at the coloured random graph models in three regimes, the
\emph{near-critical}, \emph{subcritical} and \emph{supercritical}
cases. Thus, we consider the cases when the connection probabilities
satisfy $a_n^{-1}p_n(a,b)\to C(a,b),$ for all $a,b\in\skrix,$ where
the sequence $(a_n)$ is such that either $na_n \to 1$ or $na_n \to
0$ or $na_n \to \infty$ and $C\colon\skrix\times\skrix
\to[0,\,\infty).$

The rest of the paper is organized in the following way. In
section~2 all our results are stated.  We state the Asymptotic
Equipartition Properties for both models in subsection~2.1,
beginning with the case of simple hierarchical structures and then
followed by the simple networked structures case. In subsection~2.2,
we compute the asymptotic number of bits needed to encode large
amount of data from the model of the mtDNA$^{4977}$ and the
metabolic network. Section~3 contains proof of main results. We
state and prove some large deviation principles for subcritical and
supercritical coloured random graphs. We derived our main results
from
Theorems~\ref{multgaltonwatsontree.DMS03},~\ref{main},~\ref{main1}~and~\ref{main2}.

\section{Statement of main results}\label{AEP}

\subsection{Asymptotic Equipartition Properties.}The underlying question is, how many bits are needed to store or
transmit the information contained in a structured data consisting
of $n$ units connected by  number of links?

Clearly, if no probabilistic structure is imposed, one needs of
order $n$ bits to transmit the units and of order $n^2$ bits to
transmit the links of the network data structure. By imposing a
probabilistic structure one can often transmit the structure at much
cheaper cost with arbitrarily high probability. This is explained by
the  Shannon-McMillan-Breiman theorems for networked structures
modelled as  sparse  coloured random graphs, and hierarchical
structures modelled as multitype Galton-Watson trees.

 Suppose $q$
is the distribution of a message $Y_n$  generated by a hierarchical
or networked source and let $H$ be entropy of the source. Then, we
shall say $-\log_2 q(Y_n)\approx nH$ bits with \emph{high
probability} if as $n\to\infty,$
$$-\sfrac{1}{n}\log_2 q(Y_n)\to H\, \mbox{ in probability.}$$

We denote by $\prob_n$ the law of a multitype
Galton-Watson tree  conditioned to have $n$ vertices 
and write $$ P_n(x):=\prob_n\{X=x\}, \mbox{for $x\in\skrit.$}$$ We
state the  asymptotic equipartition property (AEP) for simple
hierarchical data structures.

\begin{theorem}\label{smb.tree}
Suppose $X=(X(v):v\in V(T))$ is an irreducible, critical multitype
Galton-Watson tree with finite  type space $\skrix$ and bounded
offspring kernel ${\mathbb{Q}}$. Then, for every $\eps>0,$

$$ \lim_{n\rightarrow\infty}\prob_n\Big\{\Big|-\sfrac1n\log
P_n(X)+\sum_{(a,c)\in\skrix\times\skrix^{*}}\pi(a){\mathbb{Q}}\{c\,|\,a\}\log
{\mathbb{Q}}\{c\,|\,a\}\Big|\ge \eps \Big\}=0.$$

\end{theorem}

We can extract from Theorem~\ref{smb.tree} the following useful
information: To transmit the information contained in a large
critical multitype Galton-Watson tree  one needs with high
probability, about
$$n\Big[-\frac{1}{\log 2}\sum_{(a,c)\in\skrix\times\skrix^{*}}\pi\otimes{\mathbb{Q}}(a,c)
\log{\mathbb{Q}}\{c\,|\,a\}\Big]\qquad\mbox{bits},$$ where $n$ is
the number of vertices in the tree. We consider the following
example from the field of biology.

{\bf Mutations in mitochodrial DNA.} Mitochondria are organelles in
cells carrying their own DNA. Like nuclear DNA, mtDNA is subject to
mutations which may take the form of base substitutions, duplication
or deletions. The population mtDNA is modelled by two-type process
where the units are $a$ (normal) and $b$ (mutant), and the links are
mother-child relations. A normal can give birth to either two
normals or, if there is mutation, one normal and one mutant. Suppose
the latter happens with probability or mutation rate $\alpha.$
Mutants can only give birth to mutants. A DNA molecule may also die
without reproducing.

 Let the
survival probabilities be $\displaystyle
p\in\big[0,\,\sfrac{1}{(2-\alpha)}\big]$ and $\displaystyle
q\in\big[0,\,\sfrac{1}{2}\big]$ for normals and mutants resp. We
assume that the population is started from one normal ancestor.
Suppose the offspring kernel ${\mathbb{Q}}$ is given by
${\mathbb{Q}}\{(0,\emptyset)|a\}=1-p$,\,
${\mathbb{Q}}\{(2,(a,b))|a\}=p\alpha,$\,\,
${\mathbb{Q}}\{(2,(a,a))|a\}=p(1-\alpha),$
\,${\mathbb{Q}}\{(0,\emptyset)|b\}=1-q$ and
${\mathbb{Q}}\{(2,(b,b))|b\}=q.$ Then, the process $X$ is a
multitype Galton-Watson process with  matrix $A$ (with index set
$\{a,b\}$) given by
$$A=\begin{pmatrix}

p(2-\alpha)& 0\\
p\alpha    & 2q
\end{pmatrix}.$$

We restrict ourselves to the special case when $p=q=\sfrac12$ and
$\alpha>0.$ This case corresponds to the model for non-dividing
tissue such as the brain. This means that the population of mtDNA is
kept constant on average but that mitochondrial DNA keeps
reproducing also in non-dividing cells. See, for example Arking
 \cite{Ar98} or Olofsson and Shaw \cite{OS02} and the references therein.

 We observe that, in this
special case $X$ is critical and irreducible, with
$\pi(a)=\pi(b)=\sfrac12.$ Therefore, by Theorem~\ref{smb.tree} one
needs with high probability  approximately,
\begin{equation}\label{randomge.exa}
n\Big[1-\sfrac{1}{\log16}(\alpha\log\alpha+(1-\alpha)\log(1-\alpha))\Big]\,\mbox{
bits,}
\end{equation}
 in order to store or transmit data from a model of non-dividing
tissues. For more examples of data source with tree structure, we
refer to Kimmel and Axelrod \cite{KA02} or Mode \cite{Mo71}.\\

We state the AEP for networked data structure described by random
coloured graphs. By $\prob_n$ we also denote the (probability) law
of a coloured random graph with $n$ vertices. We write
$$P_n(x)=\prob_n\{X=x\},\mbox{ for $x\in\skrig_n.$}$$

\begin{theorem}\label{randomgsmb.sparse}
Suppose that $X$ is a  coloured random graph with colour law
$\mu\colon\skrix\rightarrow (0,1]$ and connection probabilities
$p_n$ such that $a_n^{-1}p_n(a,b) \to C(a,b)$ for some  sequence
$(a_n)$ with $a_n n \log n \to \infty$ and $\log a_n/\log n \to -1$.
Then, for every $\eps>0$,
$$\lim_{n\rightarrow\infty}\prob_n\Big\{\big|-\sfrac{1}{a_n n^{2}\log n}\log
P_n(X)-\sfrac{1}{2}\sum_{a,b\in\skrix} \mu(a) C(a,b) \mu(b) \big|\ge
\eps\Big \}=0.$$
\end{theorem}

In other words, in order to transmit a  coloured random graph in the
given regime one needs with high probability, about
$$\frac{ a_n n^{2} \log n}{2\log 2}\, \sum_{a,b\in\skrix} \mu(a) C(a,b) \mu(b)\,\,\mbox{bits.}$$
 The most interesting regime is when the cost of transmitting
colours and transmitting edges is of comparable order, i.e. when
$a_n^{-1}=n \log n.$ In this case one obtains the following
Shannon-McMillan-Breiman theorem.

\begin{theorem}\label{randomgsmb.criticals}
Suppose that $X$ is a coloured random graph with colour law
$\mu\colon\skrix\rightarrow (0,1]$ and connection probabilities
$p_n$ such that $(n \log n)\, p_n(a,b) \to C(a,b)$ for $C \colon
\skrix\times\skrix\rightarrow [0,\infty)$ symmetric. Then, for every
$\eps>0,$

$$\lim_{n\rightarrow\infty}\prob_n\Big\{\big|-\sfrac{1}{n}\log
P_n(X)- \sfrac12\sum_{a,b\in\skrix} \mu(a) C(a,b) \mu(b)
+\sum_{a\in\skrix} \mu(a) \log \mu(a)  \big|\ge \eps\Big \}=0.$$
\end{theorem}

{\bf Interpretation.} From Theorem~\ref{randomgsmb.criticals}  one
can  deduce that, the number of bits needed in order to code a
networked data structure consisting of $n$ units connected by an
average  number of order $n/\log n$ links  with high probability is
about $n H,$ where $H$ is the entropy defined by
\begin{equation}\label{randomg.aep}
H:= \frac{1}{\log2}\Big[\sfrac12 \sum_{a,b\in\skrix} \mu(a) C(a,b)
\mu(b)-\sum_{a\in\skrix} \mu(a) \log \mu(a)\Big].
\end{equation}

{\bf Metabolic network.}  We consider a metabolic network  of the
energy and biosynthesis metabolism of the bacterium E.coli. Here,
the units represent substrates and products, and  links represent
interactions.  Suppose half the nodes in the graph are of unit $a$
(substrate) and half are of unit $b$ (product), and link  between
pair of units $(a,b)$ occur independently with connection
probability $\sfrac{C(a,b)}{n},$ where
$C:\{a,\,b\}\times\{a,\,b\}\rightarrow[0,\infty)$ is nonzero
symmetric function and $n$ the size of the graph. We write
$$H:=\sfrac{1}{8\log 2} (2C(a,b)+C(a,a)+C(b,b)).$$  Then, by
Theorem~\ref{randomgsmb.sparse} one needs with high probability
about
$(n\log n)\, H\,\mbox{ bits}$
to transmit the data contained in the metabolic network of the
bacterium E.Coli.

\section{Proof of main results}
\subsection{LDP for the empirical offspring measure.} We present the recent large deviation principle for empirical
offspring measures on random trees, see Dembo \emph{et al.}
\cite{DMS03}. We recall from the introductory section that $|T|$ is
the number of vertices  and $V=V(T)$ is the set of all vertices in
tree $T$. We also recall that $m(a,c)$ is the multiplicity of the
symbol $a$ in $c=(n,a_1,\ldots,a_n)$ and that
$$\displaystyle\skrix^*=\bigcup_{n=0}^\infty \{n\} \times
\skrix^n.$$

For every multitype Galton-Watson tree $X,$ the \emph{empirical
offspring measure }$M_X$ is defined by
$$M_X(a,c)=\frac{1}{|T|}\sum_{v\in V}\delta_{(X(v),C(v))}(a,c),\,\mbox{ for
$(a,c)\in\skrix\times\skrix^{*}$}.$$ We call $\nu$
\emph{shift-invariant} if\,
$\displaystyle\nu_{1}(a)=\sum_{(b,c)\in\skrix\times\skrix^*}m(a,c)\nu(b,c),\,\mbox{
for all $a\in\skrix\times\skrix^{*}$ }.$

 We denote by $\skrim(\skrix\times\skrix^*)$ the space of
probability measures $\nu$ on $\skrix\times\skrix^*$ with $\int n \,
\nu(da \, , dc)<\infty$, using the convention
$c=(n,a_1,\ldots,a_n)$. We endow this space with the smallest
topology which makes the functionals $\nu\mapsto \int f(b,c)\,
\nu(db\, ,dc)$ continuous, for $f:\skrix\times\skrix^*\to\R$ either
bounded, or $$f(b,c)=m(a,c)\1_{b_0}(b)\,\mbox{ for some $a,b_0
\in\skrix$.}$$

\begin{theorem}[ Dembo et al. \cite{DMS03}]\label{multgaltonwatsontree.DMS03}
Suppose that $X$ is an irreducible, critical multitype Galton-Watson
tree with an offspring law whose exponential moments are all finite,
conditioned  to have exactly $n$ vertices. Then, for
$n\rightarrow\infty,$ the empirical offspring measure $M_X$
satisfies a large deviation principle in
$\skrim(\skrix\times\skrix^*)$ with speed $n$ and the convex, good
rate function
\begin{align}\label{multgaltonwatsontree.rateMX}
J(\nu)=\left\{\begin{array}{ll}H(\nu\,\|\,\nu_1\otimes{\mathbb{Q}})
& \mbox{ if  $\nu$ is shift-invariant, }\\
\infty & \mbox{otherwise.}
\end{array}\right.
\end{align}
\end{theorem}

We remark that the critical and noncritical cases give the same tree
under conditioning. See Dembo et al.  \cite{DMS03}.

\subsection{Large deviation principle for sparse random coloured
graphs.}

For any finite or countable set $\skriy$ we denote by
$\skrim(\skriy)$ the space of probability measures, and by
$\tilde\skrim(\skriy)$ the space of finite measures on $\skriy$,
both endowed with the weak topology. We denote by
$\tilde\skrim_*(\skriy \times \skriy)$ the subspace of symmetric
measures in $\tilde\skrim(\skriy \times \skriy)$. We recall that $V$
is fixed set of $n$ vertices and $E\subset\skrie:=\big\{(u,v)\in
V\times V \, : \, u<v\big\}$  is the edge set.

We associate with any coloured random graph $X$ with $n$ vertices a
probability measure, the \emph{empirical colour
measure}~$L^1\in\skrim(\skrix)$, defined ~by
$$L^{1}(a):=\frac{1}{n}\sum_{v\in V}\delta_{X(v)}(a),\quad\mbox{ for $a\in\skrix$, }$$
and a symmetric finite measure, the \emph{empirical pair measure}
$L^{2}\in\tilde\skrim_*(\skrix\times\skrix),$ defined by
$$L^{2}(a,b):=\frac{1}{n^{2}a_n}\sum_{(u,v)\in E}[\delta_{(X(v),\,X(u))}+
\delta_{(X(u),\,X(v))}](a,b),\quad\mbox{ for $a,b\in\skrix$. }$$ The
total mass $\|L^2\|$ of $L^2$ is $2|E|/(n^2a_n).$ The next theorem
is the LDP for the empirical colour measure and the empirical pair
measure of a class of sparse  coloured random graphs. i.e
 $\displaystyle a_n=\sfrac{1}{n}.$

\begin{theorem}[Doku et al. \cite{DM06a}]\label{randomg.jointL2L1}\label{main}
Suppose that $X$ is a coloured random graph with colour law $\mu$
and edge probabilities satisfying $n p_n(a,b) \to C(a,b)$ for some
symmetric function $C\colon\skrix\times\skrix\rightarrow
[0,\infty)$. Then, as $n\rightarrow\infty,$ the pair $(L^1,L^2)$
satisfies  a large deviation principle in
$\skrim(\skrix)\times\tilde{\skrim}_*(\skrix\times\skrix)$ with good
rate function,
\begin{equation}\label{randomg.rateL2L1}
I(\omega,\varpi)=H(\omega\,\|\,\mu)+\sfrac{1}{2}{\mathfrak{H}_C}(\varpi\,\|\,\omega)\,,
\end{equation}
where ${{\mathfrak H}_C}(\varpi\, \| \, \omega
):=H\big(\varpi\,\|\,C\omega\otimes\omega\big)+\|
C\omega\otimes\omega \| -\|\varpi\|\,$ is a non-negative function
and $C\omega\otimes\omega(a,b):=C(a,b)\omega(a)\omega(b).$
\end{theorem}

\begin{remark}
By exponential equivalence, see Dembo and Zeitouni
\cite[Theorem~4.2.13]{DZ98}, one can obtain from Theorem~3.2 the LDP
for $(L^1,\,L^2)$ of any coloured random graph $X$ with connection
probabilities satisfying $a_n^{-1}p_n(a,b)\to C(a,b),$ for some
sequence $(a_n)$ with $na_n\to 1$ and
$C:\skrix\times\skrix\to[0,\,\infty)$ symmetric.
\end{remark}

The proof of Theorem~\ref{randomg.jointL2L1} uses the
G\"{a}rtner-Ellis~theorem, and the technique of mixing, see
Biggins\cite{Bi04}.

\subsection{Large-deviation principles in the sub- and supercritical
cases.} We use large deviation techniques  to study asymptotic
properties of the coloured random graphs for large $n$ in the
\emph{subcritical} and \emph{supercritical} cases. In the rest of
the paper, we assume that $(a_n)\rightarrow 0$ as $n$ approaches
infinity.

\begin{theorem}\label{randomge.jointL2L1L1d}\label{main1}
Suppose that $X$ is a coloured random graph with colour law
$\mu\colon\skrix\rightarrow (0,1]$ and edge probabilities
$p_{n}:\skrix\times\skrix\rightarrow[0,1]$ satisfying
$a_{n}^{-1}p_{n}(a,b)\rightarrow C(a,b),$ for some sequence $(a_n)$
with $na_{n}\rightarrow\infty$ and
$C:\skrix\times\skrix\rightarrow[0,\infty)$ symmetric.
 Then, for  $n\rightarrow\infty,$ the pair $(L^1,L^2)$
satisfies  a large deviation principle in
$\skrim(\skrix)\times\tilde{\skrim}_{*}(\skrix\times\skrix)$ with speed
\begin{itemize}
\item[(i)]  $a_{n}n^{2}$ and  good rate function,
\begin{equation}\label{randomge.rateL2L1L1ns}
I_{1}(\omega,\varpi)=\sfrac{1}{2}{\mathfrak{H}_C}(\varpi\,\|\,\omega).
\end{equation}

\item[(ii)] $n$ and  good rate function,
\begin{align}\label{randomge.rateL2L1L1nd}
I_{2}(\omega,\varpi)=\left\{
  \begin{array}{ll}H(\,\omega\,\|\,\mu\,) & \mbox { if  $\varpi=C\omega\otimes\omega$, }\\
\infty & \mbox{otherwise.}
\end{array}\right.
\end{align}
\end{itemize}

\end{theorem}

\begin{remark}
Intuitively this means that, on the scale $a_n n^2$ the colour law
can be changed `for free', whereas on the scale $n$ once the colour
law is fixed, the edge law has to be the typical one.
\end{remark}

\begin{theorem}\label{randomge.jointL2L1L1s}\label{main2}
Suppose that $X$ is a coloured random graph with colour law
$\mu\colon\skrix\rightarrow (0,1]$ and edge probabilities
$p_{n}:\skrix\times\skrix\rightarrow[0,1]$ satisfying
$a_{n}^{-1}p_{n}(a,b)\rightarrow C(a,b),$ for some sequence $(a_n)$
with $na_{n}\rightarrow 0$ and
$C:\skrix\times\skrix\rightarrow[0,\infty)$ symmetric.
 Then, for  $n\rightarrow\infty,$ the pair $(L^1,L^2)$
satisfies  a large deviation principle in
$\skrim(\skrix)\times\tilde{\skrim}_{*}(\skrix\times\skrix)$ with speed
\begin{itemize}

\item[(i)] $a_{n}n^{2}$ and good rate function,
\begin{align}\label{randomge.rateL2L1L1ns}
I_{3}(\omega,\varpi)=\left\{
  \begin{array}{ll}\sfrac{1}{2}{\mathfrak{H}_C}(\varpi\,\|\,\omega)  & \mbox { if
  $\omega=\mu,$  }\\
\infty & \mbox{otherwise.}
\end{array}\right.
\end{align}
\item[(ii)]$n$  and  good rate function,

\begin{equation}\label{randomge.rateL2L1L1s}
I_{4}(\omega,\varpi)=H(\,\omega\,\|\,\mu\,)
\end{equation}
\end{itemize}
\end{theorem}

\begin{remark}

Intuitively this means that, on the scale $n$ the edge law can be
changed `for free',whereas on the scale $a_n n^2$ the colour law
cannot be changed.
\end{remark}

Biggins and Penman \cite{BP03} have proved large deviation principle
for the $2|E|/n(n-1)$ using the technique of mixing, see Biggins
\cite{Bi04}.\\

In the rest of the section we give the proofs of the large deviation
principles (LDPs) for coloured random graphs in the sub- and
supercritical regimes, and use our LDPs and
Theorems~\ref{multgaltonwatsontree.DMS03}~and~\ref{main}  to prove
the asymptotic equipartition properties for simple hierarchical and
networked structures. We prove our large deviation principles  using
the technique of (exponential) change of measures. Specifically, we
use the technique of change of measure to prove the Upper bounds in
Theorems~\ref{randomge.jointL2L1L1d}~and~\ref{randomge.jointL2L1L1s}.
We then obtain the proofs of all Lower bounds \emph{except}
Theorems~\ref{randomge.jointL2L1L1d}(i)~and~\ref{randomge.jointL2L1L1s}(ii)
from the Upper bounds. Lower bounds of
Theorems~\ref{randomge.jointL2L1L1d}(i)~and~\ref{randomge.jointL2L1L1s}(ii)
are proved from Lower bounds of
Theorems~\ref{randomge.jointL2L1L1d}(ii)~and~\ref{randomge.jointL2L1L1s}(i)
respectively. All our proofs use two important Lemmas, Euler's
Formula and Exponential Tightness lemma, which we shall state and
prove in  Subsection~\ref{Useful}.

\subsection{Some Useful Lemmas.}\label{Useful}

\begin{lemma}[Euler's Formula]\label{randomge.sequence1}\label{randomg.sequence1}

If\, $a_{n}^{-1}p_n(a,b)\to C(a,b),$\, for all $a,b\in\skrix$ and
$(a_n)\rightarrow 0,$   then
\begin{equation}\label{randomge.sequence}
\lim_{n\rightarrow\infty}\big[1+\alpha
p_{n}(a,b)\big]^{a_{n}^{-1}}=e^{\alpha C(a,b)},\mbox { for all
$a,b\in\skrix$ and $\alpha\in\R$.}
\end{equation}
\end{lemma}

\begin{Proof}
Observe that, for any $\eps>0$ and for large $n$  we have
\begin{equation}
\Big[1+a_{n}(\alpha C(a,b)-\eps)\Big]^{a_{n}^{-1}}\le \Big[1+ \alpha
p_{n}(a,b)\Big]^{a_{n}^{-1}}\le \Big[1+a_{n}(\alpha
C(a,b)+\eps)\Big]^{a_{n}^{-1}},
\end{equation}
by the pointwise convergence. Hence by the sandwich theorem and
Euler's formula we have (\ref{randomge.sequence}).
\end{Proof}

Note, $\prob$ is used instead of $\prob_n$ in all our large
deviation analysis for sake of a simple presentation.

\begin{lemma}[Exponential Tightness]\label{randomge.tightness}\label{randomg.tightness}
For every $\alpha>0,$ \, there exists $N\in \N$ such that
\begin{equation}\label{randomge.qtightness}
\limsup_{n\rightarrow\infty}\sfrac{1}{a_{n}n^{2}}\log\prob\Big\{|E|>a_n
 n^{2}N \Big\}\le-\alpha.
\end{equation}
\end{lemma}

\begin{Proof}
Let $c>\max_{a,b\in\skrix} C(a,b)>0.$ Using Chebysheff's inequality
and Lemma~\ref{randomge.sequence1}, we have (for sufficiently
large~$n$)
\begin{align*}
\prob\Big\{ |E|\ge a_n n^{2} l\Big\} &\le e^{-a_n n^{2}l}\me\Big\{e^{|E|}\Big\}\\
 &\le e^{-a_n n^{2}l}\sum_{k=0}^{n(n-1)/2}e^{k} \left(\heap{n(n-1)/2}{
 k}\right)\Big(a_nc)  \Big)^{k}\Big(1-a_nc \Big)^{n(n-1)/2-k}\\
&= e^{-a_n n^{2}l}\Big( 1+(e-1)a_nc\Big)^{a_n^{-1}(a_n n(n-1)/2)}\\
&\le e^{-a_n n^{2}l}e^{a_n n^{2}(c(e-1+o(1)))}.
\end{align*}

Now given $\alpha$ choose $N\in\N$  such that $N>\alpha+c(e-1)$ and
observe that, for sufficiently large $n,$
\begin{equation}\nonumber
\prob\Big\{|E|\ge a_n n^{2} N\Big\}\le e^{-a_n n^{2}\alpha},
\end{equation}
which implies the statement.
\end{Proof}

\subsection{Change of measure on the scale $n$.}

We denote by $\skric_2$  the space of symmetric functions on
$\skrix\times\skrix$ and by $\skric_1$ the space of functions on
$\skrix.\,\,$

Given a function $\tilde{f}\colon\skrix\rightarrow\R$ and  a
symmetric function $\tilde{g}\colon
\skrix\times\skrix\rightarrow\R,$ define the constant
$U_{\tilde{f}}$ by
$$U_{\tilde{f}}=\log\sum_{a\in\skrix}e^{\tilde{f}(a)}\mu(a).$$
For the function $\tilde{g}\in\skric_2$ we define the symmetric
function $\tilde{h}_n^{(2)}:\skrix\times\skrix\rightarrow\R$ by

\begin{equation}\label{randomge.tildeh2super}
\tilde{h}_n^{(2)}(a,b)=\log\Big[\Big(1-p_{n}(a,b)+p_{n}(a,b)e^{\tilde{g}(a,b)/na_{n}}\Big)^{-n}\Big].\\
\end{equation}

 Use $\tilde{f}$,\, $\tilde{g}$ to
define (for sufficiently large $n$) a new coloured random graph  in
the following manner.
\begin{itemize}
 \item To the  vertices  $V=\{\,1,\,\ldots,\, n\,\}$ we assign colours from
$\skrix$ independently and identically according to the colour law
$\tmu$ defined by
$$ \tmu(a)=e^{\tilde{f}(a)-U_{\tilde{f}}}\mu(a).$$

\item Given  any two vertices $u,v\in V,$ with $u$ carrying colour $a$ and
$v$ carrying colour $b$  connect vertex  $u$ to vertex $v$ with
probability
\begin{equation}\label{randomg.cfm1}
\tilde{p}_{n}(a,b)=\sfrac{p_{n}(a,b)e^{\tilde{g}(a,b)/na_{n}}}{1-p_{n}(a,b)+p_{n}(a,b)
e^{\tilde{g}(a,b)/na_{n}}},
\end{equation}
 otherwise keep $u$ and $v$ disconnected.
\end{itemize}

For this new graph, observe  $\tmu$ is a probability measure and
further that $\tilde{p}_n(a,\,b)\in[0,\,1],\,$ for all
$a,b\in\skrix.$ Denote by   $\tilde{\P}$  the law of the new
 coloured random graph construct from  $\tmu$ and  $\tilde{p}.$ We
note from the construction of the new graph that  $\tilde{\P}$ is
absolutely continuous with respect to $\P,$ as for a coloured random
graph $X,$

\begin{align}
\frac{d\tilde{\P}}{d\P}(X) &=\prod_{u\in V}\sfrac{\tmu(X(u))}{\mu(X(u))}\prod_{(u,v)\in E}\sfrac{\tilde{p}_{n}(X(u),X(v))}{p_{n}(X(u),X(v))}\prod_{(v,u)\not\in E}\sfrac{1-\tilde{p}_{n}(X(u),X(v))}{1-p_{n}(X(u),X(v))}\nonumber\\
&=\prod_{u\in V}\sfrac{\tmu(X(u))}{\mu(X(u))}\prod_{(u,v)\in E}\sfrac{\tilde{p}_{n}(X(u),X(v))}{p_{n}(X(u),X(v))}\times\sfrac{n-np_{n}(X(u),X(v))}{n-n\tilde{p}_{n}(X(u),X(v))}\prod_{(u,v)\in\skrie}\sfrac{n-n\tilde{p}_{n}(X(u),X(v))}{n-np_{n}(X(u),X(v))}\nonumber\\
&=\prod_{u\in V}e^{\tilde{f}(X(u))-U_{\tilde{f}}}\prod_{(u,v)\in E}e^{\tilde{g}(X(u),X(v))/na_n}\prod_{(u,v)\in\skrie}e^{\tilde{h}_n^{(2)}(X(u),X(v))/n}\nonumber\\
&=e^{ n\,\langle L^1,\,
\tilde{f}-U_{\tilde{f}}\rangle+n\,\langle\sfrac{1}{2}L^2,\,\tilde{g}\rangle+n\,\langle\sfrac{1}{2}L^1\otimes
L^1,\,\tilde{h}_{n}^{(2)}\rangle-\,\langle\,\sfrac{1}{2}L_{\Delta}^{1},\,\tilde{h}_n^{(2)}\rangle},
\label{randomge.Itransform3super}
\end{align}
where $L_{\Delta}^{1}(a,\,a) =\sfrac{1}{n}\sum_{u\in
V}\delta_{(X(u),X(u))}(a,\,a),$ for $a\in\skrix$  and
$\sum_{a\in\skrix}L_{\Delta}^{1}(a,\,a)=1.$

\subsection{Change of measure on the scale $a_nn^2.$} Define for  $\tilde{g}\in\skric_2$, $\tilde{h}_n^{(1)}:\skrix\times\skrix\rightarrow\R$ by

$$
\tilde{h}_n^{(1)}(a,b)=-\log\Big[\Big(1-p_{n}(a,b)+p_{n}(a,b)e^{\tilde{g}(a,b)}\Big)^{1/a_n}\Big].
$$

Define for $\tilde{f}\in\skric_1$  and $\,\tilde{g}$  a new coloured
random graph (for sufficiently large $n$) in the following way:

\begin{itemize}
 \item Assign to the $n$ vertices in $V$ colours from
$\skrix$ independently and identically according to the colour law
$\tmu$ defined by
\begin{equation}\label{randomg.mcsup}
\tmu(a)=e^{\tilde{f}(a)-U_{\tilde{f}}}\mu(a).
\end{equation}

\item Given  any two vertices $u,v\in V,$ with $u$ carrying colour $a$ and
$v$ carrying  colour $b$  connect vertex  $u$ to vertex $v$ with
probability
\begin{equation}\label{randomg.cfmsup}
\tilde{p}_{n}(a,b)=\sfrac{p_{n}(a,b)e^{\tilde{g}(a,b)}}{1-p_{n}(a,b)+p_{n}(a,b)
e^{\tilde{g}(a,b)}},
\end{equation}
 otherwise keep $u$ and $v$ disconnected.
\end{itemize}

 Note the colour law $\tilde{\mu}$ is a
 probability measure and  the connection probabilities  satisfy  $\tilde{p}_n(a,\,b)\in[0,\,1],
 $\,\,for all $ a,b\in\skrix.$ We denote by $\tilde{\P}$ the law of the  coloured random graph
obtained from $\tmu$ and $\tilde{p}_n.$ By construction $\tilde{\P}$
is absolutely
 continuous with respect to $\P,$ as for a coloured random graph $X,$
\begin{align}
\frac{d\tilde{\P}}{d\P}(X) &=\prod_{u\in
V}\sfrac{\tmu(X(u))}{\mu(X(u))}
\prod_{(u,v)\in E}\sfrac{\tilde{p}_{n}(X(u),X(v))}{p_{n}(X(u),X(v))}\prod_{(v,u)\not\in E}\sfrac{1-\tilde{p}_{n}(X(u),X(v))}{1-p_{n}(X(u),X(v))}\nonumber\\
&=\prod_{u\in V}e^{\tilde{f}(X(u))-U_{\tilde{f}}}\prod_{(u,v)\in
E}\sfrac{\tilde{p}_{n}(X(u),X(v))}{p_{n}(X(u),X(v))}\times\sfrac{1-\tilde{p}_{n}(X(u),X(v))}{1-p_{n}(X(u),X(v))}
\prod_{(u,v)\in\skrie}\sfrac{1-p_{n}(X(u),X(v))}{1-\tilde{p}_{n}(X(u),X(v))}\nonumber\\
&=\prod_{u\in V}e^{\tilde{f}(X(u))-U_{\tilde{f}}}\prod_{(u,v)\in
E}e^{\tilde{g}(X(u),\,X(v))}
\prod_{(u,v)\in\skrie}e^{a_n\tilde{h}_n^{(1)}(X(u),\,X(v))}\nonumber\\
&=e^{n\,\langle L^1,\,
\tilde{f}-U_{\tilde{f}}\rangle+a_{n}n^2\,\langle\sfrac{1}{2}L^2,\,\tilde{g}\rangle+a_{n}n^{2}\,\langle\sfrac{1}{2}L^1\otimes
L^1,\,\tilde{h}_n^{(1)}\rangle-a_{n}n^{2}\,\langle\sfrac{1}{2}L_{\Delta}^{2},\,\tilde
{h}_n^{(1)}\rangle},\label{randomge.Itransform1sup}
\end{align}
where $L_{\Delta}^{2}(a,\,a) =\sfrac{1}{n^2}\sum_{u\in
V}\delta_{(X(u),X(u))}(a,\,a),$ for $a\in\skrix$  and
$\sum_{a\in\skrix}L_{\Delta}^{2}(a,\,a)=\sfrac{1}{n}.$

\subsection{Upper bound in
Theorem~\ref{randomge.jointL2L1L1d}(ii).}\label{LDPsn}
 Define for
$(\omega,\varpi)\in\skrim(\skrix)\times\tilde{\skrim}_{*}(\skrix\times\skrix),$
$\hat{I}_{2}(\omega,\,\varpi)$ by
\begin{equation}
\hat{I}_{2}(\omega,\,\varpi)=\sup_{\heap{\tilde{f}\in\skric_1}{\tilde{g}\in\skric_{2}}}\Big\{\sum_{a\in\skrix}\big(\tilde{f}(a)-U_{\tilde{f}}\big)\omega(a)
+\sum_{a,b\in\skrix}\sfrac{1}{2}\tilde{g}(a,b)(\varpi(a,b)-C(a,b)\omega(a)\omega(b))\Big\}\\
\label{randomge.uppboundnsuper}
\end{equation}

\begin{lemma}\label{randomge.L2L1uppboundnsuper}
 For each closed set
$F\subset\tilde{\skrim}_{*}(\skrix\times\skrix),$ we have
\begin{equation}
\limsup_{n\rightarrow\infty}\sfrac{1}{n}\log\P\big\{(L^1,\,L^2)\in
F\big\}\le-\inf_{(\omega,\,\varpi)\in
F}\hat{I}_{2}(\omega,\,\varpi).\nonumber
\end{equation}
\end{lemma}
\begin{Proof}
Fix $\tilde{f}\in\skric_1.$ For any  $\tilde{g}\in\skric_2,$  we
define $\tilde{\beta}\colon\skrix\times\skrix\rightarrow\R$ by
$\displaystyle\tilde{\beta}(a,b)=-\tilde{g}(a,b)C(a,b).$

We notice from Lemma~\ref{randomge.sequence1} that, $\displaystyle
\lim_{n\rightarrow\infty}\tilde{h}_n^{(2)}(a,b)=\tilde{\beta}(a,b),$\,
for all $a,b\in\skrix.$  Hence, for  any  $\delta>0$ and for
(sufficiently) large $n,$ we have
\begin{equation}\label{esiamp.equ1}
\tilde{h}_n^{(2)}(a,b)\le |\tilde{\beta}(a,b)|+\delta,\,\mbox{ for
all $a,b\in\skrix.$}
\end{equation}

Using (\ref{randomge.Itransform3super}) and \eqref{esiamp.equ1} we
obtain
\begin{equation}\nonumber
e^{(\max_{a\in\skrix} |\tilde{\beta}(a,a)| +\delta)/2}\ge \int
e^{\,\langle\frac{1}{2}L_{\Delta}^{1},\,
\tilde{h}_n^{(2)}\rangle}d\tilde{\prob} =\me\Big\{e^{n\,\langle
L^1,\, \tilde{f}-U_{\tilde{f}}\rangle+n\,\langle\frac{1}{2}L^2,\,
\tilde{g}\rangle+n\,\langle\frac{1}{2}L^1\otimes L^1,\,
\tilde{h}_n^{(2)}\rangle}\Big\},
\end{equation}
for any $\delta>0$ and for large $n.$ Therefore, we have
\begin{equation}\label{randomg.estPnsuper}
\limsup_{n\rightarrow\infty}\sfrac{1}{n}\log\me\Big\{e^{n\,\langle
\,L^1,\,\tilde{f}-U_{\tilde{f}}\rangle+n\,\langle
\frac{1}{2}\,L^2,\,
\tilde{g}\rangle+n\,\langle\frac{1}{2}\,L^1\otimes L^1,\,
\tilde{h}_n^{(2)}\rangle}\Big\}\le 0.
\end{equation}

We now fix $\eps>0$ and write
$\hat{I}_{2}^{\eps}(\omega,\,\varpi):=\min\{\hat{I}_{2}(\omega,\,\varpi),{\eps}^{-1}\}-\eps.$
Let $F$ be a closed subset of
$\tilde{\skrim}_{*}(\skrix\times\skrix)$  and suppose
$(\omega,\,\varpi)\in F.$  Choose
$\tilde{f}\in\skric_1$,\,\,$\tilde{g}\in\skric_2$ such that\,\,
$$\displaystyle\langle\,\omega,\,\tilde{f}-U_{\tilde{f}}\rangle+\sfrac{1}{2}\,\langle\,
\varpi,\,\tilde{g}\rangle-\sfrac{1}{2}\,\langle\,\omega\otimes\omega,\,
C\tilde{g}\rangle\ge\hat{I}_{2}^{\eps}(\omega,\,\varpi).$$

Since $\skrix$ is finite,  we can find  open neighbourhoods
$B_{\varpi}^{2}$ and $B_{\omega}^{1}$ of $\varpi,\,\omega$ such that
\begin{equation}\nonumber
\inf_{\tilde{\omega}\in B_{\omega}^{1},\,\tilde{\varpi}\in
B_{\varpi}^{2}}\big\{\langle\tome,\, \tilde{f}-U_{\tilde{f}}
\rangle+ \sfrac12 \, \langle\,\tilde\varpi,\,
\tilde{g}\rangle-\sfrac 12 \, \langle\,\tome\otimes\tome,\,
C\tilde{g}\rangle\big\}\ge \hat{I}_{2}^{\eps}(\omega,\varpi)-\eps.
\end{equation}

Using  Chebysheff's inequality  and (\ref{randomg.estPnsuper}), we
have that

\begin{equation}\begin{aligned}
\limsup_{n\rightarrow\infty}\sfrac{1}{n}& \log\prob\big\{(L^1,
L^2)\in
B_{\omega}^{1}\times B_{\varpi}^{2}\big\}\\
&\le\limsup_{n\rightarrow\infty}\sfrac{1}{n}\log
\me\Big\{e^{n\,\langle
L^1,\,\tilde{f}-U_{\tilde{f}}\rangle+n\,\langle
\frac{1}{2}\,L^2,\,\tilde{g}\rangle+n\,\langle\frac{1}{2}L^1\otimes
L^1,\,\tilde{h}_n^{(2)}\rangle}\Big\}-\hat{I}_2^{\eps}(\omega,\varpi)+\eps\\
&\le
-\hat{I}_{2}^{\eps}(\omega,\varpi)+\eps.\label{randomg.LDPballssuper}
\end{aligned}\end{equation}

Use  Lemma~\ref{randomge.tightness} to choose $N\in\N$ such that
\begin{equation}\nonumber
\limsup_{n\rightarrow\infty}\sfrac{1}{n}\log\P \Big\{|E|>a_{n}n^{2}
N \Big\}=-\infty.
\end{equation}

For this $N$ define the set $K_{N}$ by
$$\displaystyle
K_{N}=\big\{(\omega,\,\varpi)\in\skrim(\skrix)\times\tilde{\skrim}_{*}(\skrix\times\skrix):\|\varpi\|\le
2N\big\}.$$

The set $K_{N}\cap F$ is compact and therefore may be covered by
finitely many sets

$B_{\omega_ {r}}^{1}\times B_{\varpi_ {r}}^{2} \mbox{ with $
(\omega_{r},\varpi_{r})\in F$ for $r=1,\ldots ,m.$}$ Hence, we have
\begin{equation}\nonumber
\prob\big\{(L^1,\, L^2)\in F\big\}\le\sum_{r=1}^{m}\prob\big\{(L^1,
L^2)\in B_{\omega_{r}}^{1}\times
B_{\varpi_{r}}^{2}\big\}+\prob\big\{(L^1, L^2)\not\in K_{N}\big\}.
\end{equation}

We may now use (\ref{randomg.LDPballssuper})  to obtain, for all
sufficiently small $\eps>0$,
\begin{equation}\nonumber
\begin{aligned}
\limsup_{n\rightarrow\infty}\sfrac{1}{n}\log\prob\big\{(L^1, L^2)\in
F\big\}& \le
\max_{r=1}^{m}\limsup_{n\rightarrow\infty}\sfrac{1}{n}\log\prob\big\{(L^1,
L^2)\in
B_{\omega_{r}}^{1}\times B_{\varpi_{r}}^{2}\big\}-\infty \\
& \le -\inf_{(\omega,\,\varpi)\in
F}\hat{I}_{2}^{\eps}(\omega,\,\varpi)+\eps.
\end{aligned}
\end{equation}
Taking $\eps\downarrow 0$ we have the desired statement.
\end{Proof}

We solve the variational problem on the right side of equation
(\ref{randomge.uppboundnsuper}).\\

\begin{lemma}\label{randomge.Vratensuper}

 $\hat{I}_{2}(\omega,\,\varpi)=H(\omega\,\|\,\mu)$  if \emph{(}and only if\emph{)}
 $\varpi=C\omega\otimes\omega,$
and $\infty$ otherwise.

\end{lemma}
\begin{Proof}
Suppose that $\varpi\not =C\omega\otimes\omega.$ Then there exists
$a_0,b_0\in\skrix$ such $\varpi(a_0,b_0)>
C(a_0,b_0)\omega(a_0)\omega(b_0)$ or $\varpi(a_0,b_0)<
C(a_0,b_0)\omega(a_0)\omega(b_0).$   Define for this
$a_0,b_0\in\skrix$ the symmetric function $\tilde{g}$ by
\begin{equation}\label{randome.W}
\tilde{g}(a,b)=K(\1_{(a_0,\,b_0)}(a,b)+\1_{(b_0,\,a_0)})(a,b),\,\mbox{for
$a,b\in\skrix$ and $K\in\R$ }
\end{equation}
 Considering this $\tilde{g}$ in
\eqref{randomge.uppboundnsuper} we have

\begin{equation}\label{esiamp.equ2}
\begin{aligned}
\sum_{a,b\in\skrix}
\sfrac{1}{2}\tilde{g}(a,b)\varpi(a,b)+\sum_{a,b\in\skrix}
&-\sfrac{1}{2}\tilde{g}(a,b)C(a,b)\omega(a)\omega(b)\\
&=K(\varpi(a_0,b_0)-C(a_0,b_0)\omega(a_0)\omega(b))
\xrightarrow{|K|\uparrow\infty}\infty,
\end{aligned}
\end{equation}
where the sign of  $|K|$ is chosen such that expression in the right
side of \eqref{esiamp.equ2} remain positive.

Suppose that $\varpi= C\omega\otimes\omega.$ Then, by the
variational characterization of the relative entropy we have
$I_2(\omega,\, \varpi)=H(\omega\,\|\,\mu),$ which ends the proof of
the upper bounds.
\end{Proof}

\subsection{Lower bound Theorem~\ref{randomge.jointL2L1L1d}(ii).}

\begin{lemma}\label{randomg.losnsuper}
For every open set $
O\subset\skrim(\skrix)\times\tilde{\skrim}_{*}(\skrix\times\skrix),$
\begin{equation}
\liminf_{n\rightarrow\infty}\sfrac{1}{n}\log\P\big\{(L^1,L^2)\in
O\big\}\ge -\inf_{(\omega,\,\varpi)\in O} I_{2}(\omega,\,\varpi).
\end{equation}
\end{lemma}
\begin{Proof}
Suppose $(\omega,\,\varpi)\in O$ is such that we have
$\varpi=C\omega\otimes\omega.$ Set $\tilde{g}(a,b)=0,$  for all
$a,b\in\skrix$ and define $\tilde{f}_{\omega}\colon\skrix\rightarrow
\R$ by
\begin{equation}\label{randomg.S1super}
\begin{aligned}\nonumber
\tilde{f}_{\omega}(a)=\left\{\begin{array}{ll}\log\sfrac{\omega(a)}{\mu(a)},
&\mbox{if $\omega(a)> 0$,  }\\
0, & \mbox{otherwise.}
\end{array}\right.
\end{aligned}
\end{equation}

We note that this choice of $\tilde{g}$  yields
$\tilde{h}_{n}^{(2)}(a,b)=0,\,$ for all $a,b\in\skrix.$ Choose
$B_{\omega}^{1},B_{\varpi}^{2}$ open neighbourhoods of
$\omega,\varpi,$ such that $B_{\omega}^{1}\times
B_{\varpi}^{2}\subset O$ and for all
$(\tilde{\omega},\tilde{\varpi})\in B_{\omega}^{1}\times
B_{\varpi}^{2},$ $$\langle \tilde{f}_{\omega},\omega\rangle-\eps\le
\langle \tilde{f}_{\omega},\tome\rangle.$$

We use the  probability measure $\tilde{\P}$ given by
$\tilde{g}_{\varpi}.$  We observe that the colour law is $\omega$
and the connection probabilities satisfy
$a_n^{-1}\tilde{p}_n(a,b)\to
\tilde{C}(a,b):=\varpi(a,b)/(\omega(a)\omega(b)),$ as $n$ approaches
infinity. Therefore, using (\ref{randomge.Itransform3super}) we have
that
\begin{equation}\nonumber
\begin{aligned}
\prob\Big\{(L^1,L^2) \in
O\Big\}\ge\tilde{\me}\Big\{\sfrac{d\prob}{d\tilde{\prob}}(X)\one_{\{(L^1,L^2)\in
B_{\omega}^{1}\times B_{\varpi}^{2}\}}\Big\}
&=\tilde{\me}\Big\{ e^{-n\langle
L^1,\tilde{f}_{\omega}\rangle}\one_{\{(L^1,\,L^2)\in
B_{\omega}^{1}\times B_{\varpi}^{2}\}}\Big\}\\
 &\ge
e^{-n\langle\,\omega,\,\tilde{f}_{\omega}\rangle-n\eps}\times\tilde{\prob}\Big\{(L^1,L^2)\in
B_{\omega}^{1}\times B_{\varpi}^{2}\Big\}.
\end{aligned}
\end{equation}

Therefore, we have
\begin{equation}\label{randomg.lowbound2super}
\begin{aligned}
\liminf_{n\rightarrow\infty}\sfrac{1}{n}\log\prob\big\{(L^1,L^2)\in
O\big\} \ge -\langle\,\omega,\, \tilde{f}_{\omega}\rangle-\eps
+\liminf_{n\rightarrow\infty}\sfrac{1}{n}\log\tilde{\prob}
\big\{(L^1,L^2)\in B_{\omega}^{1}\times
B_{\varpi}^{2}\big\}.\nonumber
\end{aligned}
\end{equation}

The result follows once we prove that
\begin{equation}\label{randomg.translowboundnsuper}
\liminf_{n\rightarrow\infty}\sfrac{1}{n}\log\tilde{\prob}\big\{(L^1,L^2)\in
B_{\omega}^{1}\times B_{\varpi}^{2}\big\}= 0.
\end{equation}

We  use the upper bound (but now with the law  $\P$ replaced by
$\tilde{\P}$) to prove (\ref{randomg.translowboundnsuper}).\\

Therefore, we have

\begin{align*}
\limsup_{n\rightarrow\infty}\sfrac{1}{n}\log\tilde\prob\big\{(L^1,L^2)
\in (B_{\omega}^{1}\times B_{\varpi}^{2})^{c} \big\}&\le
-\inf_{(\tome,\tilde{\varpi})\in \tilde{F}}
\tilde{I}_{2}(\tome,\tilde\varpi),
\end{align*}

\begin{align}\nonumber
\tilde{I}_{2}(\tome,\tilde{\varpi})=\left\{
  \begin{array}{ll}H(\,\tome\,\|\,\omega\,) & \mbox { if  $\tilde{\varpi}=\tilde{C}\tome\otimes\tome$, }\\
\infty & \mbox{otherwise,}
\end{array}\right.
\end{align}

where $\tilde{F}=(B_{\omega}^{1}\times B_{\varpi}^{2})^{c}.$ It
therefore suffices to show that the infimum is positive. Suppose for
contradiction that there exists a sequence
$(\tome_n,\tilde{\varpi}_n)\in\tilde{F}$ with
$\tilde{I}_{2}(\tome_n,\tilde\varpi_n)\downarrow 0.$ Then, since
$\tilde{I}_2$ is a good rate function and its level sets are
compact, and  the mapping $({\tome},\tilde{\varpi})\mapsto
\tilde{I}_{2}({\tome},\tilde{\varpi})$ is  lower semicontinuous, we
can construct a limit point $(\tome,\tilde\varpi)\in\tilde{F}$ with
$\tilde{I}_{2}(\tome,\tilde{\varpi})=0.$  By
Lemma~\ref{randomge.Vratensuper} this implies
$H(\tome\,\|\,\omega)=0$ and $\tilde{\varpi}=C\tome\otimes\tome$,
hence $\tome=\omega,$ and
$\tilde{\varpi}=\tilde{C}\tome\otimes\tome=\varpi.$ This contradicts
$(\tome,\tilde{\varpi})\in\tilde{F}$.

\end{Proof}

\subsection{ Upper bound in Theorem~\ref{randomge.jointL2L1L1d}(i).}
We  obtain the upper bound in a variational formulation. We define
for
  $(\omega,\varpi)\in\skrim(\skrix)\times\tilde{\skrim}_{*}(\skrix\times\skrix)$ the rate function $\hat{I}_{1}$ by

\begin{equation}
\begin{aligned}
\hat{I}_{1}(\omega,\varpi)=\sup_{\tilde{g}\in\skric_{2}}\Big\{\sum_{a,b\in\skrix}\sfrac{1}{2}
\tilde{g}(a,b)\varpi(a,b)+\sum_{a,b\in\skrix}\sfrac{1}{2}(1-e^{\tilde{g}(a,b)})
C(a,b)\omega(a)\omega(b)\Big\}.\label{randomge.uppbounddsup}
\end{aligned}
\end{equation}

\begin{lemma}\label{randomge.L2L1uppboundsup} For each closed set
$F\subset\skrim(\skrix)\times\tilde{\skrim}_{*}(\skrix\times\skrix),$\,
we have
$$\qquad\limsup_{n\rightarrow\infty}\sfrac{1}{a_{n}n^{2}}\log\P\big\{(L^1,\,L^2)\in
F\big\}\le-\inf_{(\omega,\,\varpi)\in
F}\hat{I}_{1}(\omega,\,\varpi).$$
\end{lemma}
\begin{Proof}
For any  $\tilde{g}\in\skric_2$ we define $
\tilde{\beta}\colon\skrix\times\skrix\rightarrow\R$ by
$$\tilde{\beta}(a,b)=(1-e^{\tilde{g}(a,b)})C(a,b).$$  From
Lemma~\ref{randomge.sequence1} we note that
$\displaystyle\lim_{n\rightarrow\infty}\tilde{h}_n^{(1)}(a,b)=\tilde{\beta}(a,b),$
for all $a,b\in\skrix.$ We observe that, for any $\delta>0$ and for
(sufficiently) large $n$ we have
\begin{equation}\label{esiamp.equ2}
\tilde{h}_n^{(2)}(a,b)\le |\tilde{\beta}(a,b)|+\delta,\,\mbox{ for
all $a,b\in\skrix.$}
\end{equation}

We take $\tilde{f}(a)=0,\,$ for all $a\in\skrix,$ and  use
(\ref{randomge.Itransform1sup}) and \eqref{esiamp.equ2} to obtain
\begin{equation}\nonumber
\begin{aligned}
e^{na_n(\, \max_{a\in\skrix}|\tilde{\beta}(a,a)|+\delta)/2}\ge\int
e^{a_{n}n^{2}\,\langle
\sfrac{1}{2}L_{\Delta}^{2},\,\tilde{h}_n^{(1)}\rangle}d\tilde{\P}=\me\Big\{e^{a_{n}n^2\,\langle\sfrac{1}{2}\,
L^2,\,\tilde{g}\rangle+a_{n}n^{2}\,\langle\sfrac{1}{2}\,L^1\otimes
L^1,\,\tilde{h}_n^{(1)}\rangle}\Big\},
\end{aligned}
\end{equation}

for any $\delta>0$ and for  large $n.$ Therefore, we have that
\begin{equation}\label{randomge.estPsup}
\limsup_{n\rightarrow\infty}\sfrac{1}{a_{n}n^{2}}\log\me\Big\{e^{a_{n}n^2\,\langle\sfrac{1}{2}
L^2,\,\tilde{g}\rangle+a_{n}n^{2}\,\langle\sfrac{1}{2}L^1\otimes
L^1,\,\tilde{h}_n^{(1)}\rangle}\Big\}\le 0.
\end{equation}

Fix $\eps>0$ and take
$\hat{I}_{1}^{\eps}(\omega,\,\varpi)=\min\{\hat{I}_{1}(\omega,\,\varpi),{\eps}^{-1}\}-\eps.$
Suppose $(\omega,\,\varpi)\in F$  and choose $\tilde{g}\in\skric_2$
such that $$\displaystyle\sfrac{1}{2}\langle
\varpi,\,\tilde{g}\rangle+\sfrac{1}{2}\langle
\omega\otimes\omega,\,\tilde{\beta}\rangle
\ge\hat{I}_1^{\eps}(\omega,\varpi).\,$$ Using the finiteness of
$\skrix$ we can find open neighbourhoods $B_{\omega}^{1},$
$B_{\varpi}^{2}$ of $\omega,\,\varpi$ such that
\begin{equation}\nonumber
\inf_{\tilde{\omega}\in B_{\omega}^{1},\,\tilde{\varpi}\in
B_{\varpi}^{2}}\big\{\sfrac12  \langle\tilde\varpi,\,
\tilde{g}\rangle+\sfrac 12\langle\tome\otimes\tome,\,
\tilde{\beta}\rangle\big\}\ge\hat{I}_{1}^{\eps}(\omega,\,\varpi)-\eps.
\end{equation}

By Chebysheff's inequality  and (\ref{randomge.estPsup}), we have
that
\begin{align}
\limsup_{n\rightarrow\infty}\sfrac{1}{a_{n}n^{2}}\log\P\big\{(L^1,\,&
L^2)
\in B_{\omega}^{1}\times B_{\varpi}^{2}\big\}\nonumber\\
&\le\limsup_{n\rightarrow\infty}\sfrac{1}{a_{n}n^{2}}\log
\me\Big\{e^{a_{n}n^2\,\langle\sfrac{1}{2}
L^2,\,\tilde{g}\rangle+{a_{n}}n^{2}\,\langle\sfrac{1}{2}L^1\otimes
L^1 ,\,\tilde{h}_n^{(1)}\rangle}\Big\}
-\hat{I}_{1}^{\eps}(\omega,\,\varpi)+\eps\nonumber\\
&\le-\hat{I}_{1}^{\eps}(\omega,\,\varpi)+\eps.\label{randomge.LDPballssup}
\end{align}
We use Lemma~\eqref{randomge.tightness}  with $\alpha=\eps^{-1}$ to
choose $N(\eps)\in\N$
 such that
\begin{equation}\nonumber
\limsup_{n\rightarrow\infty}\sfrac{1}{a_{n}n^{2}}\log\P\Big\{|E|>a_{n}n^{2}N(\eps)
\Big\}\le-\eps^{-1}.
\end{equation}
Define for this $N,$   the set $K_{N(\eps)}$ by $$\displaystyle
K_{N(\eps)}=\big\{(\omega,\,\varpi)\in\skrim(\skrix)\times\tilde{\skrim}_{*}(\skrix\times\skrix):\|\varpi\|\le
2N(\eps)\big\}.$$

 Now, observe that
$K_{N(\eps)}\cap F$ is compact and therefore may be covered by
finitely many sets $B_{\omega}^{1}\times B_{\varpi_ {r}}^{2},$
$r=1,...,m$ with $ (\omega_{r},\,\varpi_{r})\in F$ for $r=1,...,m.$
Hence, we have
\begin{equation}\nonumber
\P\big\{ (L^1,\, L^2)\in F\big\}\le\sum_{r=1}^{m}\P\big\{(L^1,\,
L^2)\in B_{\omega_r}^{1}\times B_{\varpi_r}^{2}\big\}+\P\big\{
(L^1,\,L^2)\not\in K_{N(\eps)}\big\}.
\end{equation}
Using (\ref{randomge.LDPballssup})  for small enough $\eps>0,$ we
obtain

\begin{equation}\nonumber
\begin{aligned}
\limsup_{n\rightarrow\infty}\sfrac{1}{a_{n}n^{2}}\log\P\big\{(L^1,\,
L^2)\in F\big\}&\le
\max_{r=1}^{m}\limsup_{n\rightarrow\infty}\sfrac{1}{a_{n}n^{2}}\log\P\big\{
(L^1,\, L^2)\in B_{\omega_r}^{1}\times
B_{\varpi_r}^{2}\big\}-\eps^{-1}\\
& \le -\hat{I}_{1}^{\eps}(\omega,\,\varpi)+\eps.
\end{aligned}
\end{equation}
Taking $\eps\downarrow 0$ we have the desired statement.
\end{Proof}

We identify the rate function by solving the variational problem in
the right side of equation (\ref{randomge.uppbounddsup}).

\begin{lemma}\label{randomge.Vratesup}
 For any
$(\omega,\,\varpi)\in\skrim(\skrix)\times\tilde{\skrim}_{*}(\skrix\times\skrix)$
we have
 $\hat{I}_{1}(\omega,\,\varpi)= {\mathfrak{H}_C}(\varpi\,\|\,\omega).$

\end{lemma}
\begin{Proof} (i) Suppose  $\varpi\not\ll C\omega\otimes\omega\,.$ Then there exists
$a_{0},\,b_{0}\in\skrix$ with
$C(a_{0},b_{0})\omega(a_{0})\omega(b_{0})=0$ and
$\varpi(a_{0},b_{0})>0.$  For this $(a_{0},b_{0})$ we define the
symmetric function $\tilde{g}$ by
\begin{equation}\nonumber
\tilde{g}(a,b)=\log(K(\1_{(a_{0},b_{0})}(a,b)+\1_{(b_{0},a_{0})}(a,b))+1),
\end{equation}

for $a,b\in\skrix$ and $K>0.$ Considering our $\tilde{g}$ in
\eqref{randomge.uppbounddsup} we have
\begin{equation}\nonumber
\begin{aligned}
\sum_{a,b\in\skrix}
\sfrac{1}{2}\tilde{g}(a,b)\varpi(a,b)+\sum_{a,b\in\skrix}\sfrac{1}{2}(1-e^{\tilde{g}(a,b)})C(a,b)\omega(a)\omega(b)
=\log( K+1)(\varpi(a_{0},b_{0}))\xrightarrow{
K\uparrow\infty}\infty.
\end{aligned}
\end{equation}
Suppose that  $\varpi\ll C\omega\otimes\omega.$ Then, we have
\begin{equation}\label{randomg.equu}
\begin{aligned}
\hat{I}(\omega,\varpi) \ge\sfrac{1}{2} \sup_{g\in\skric_{2}}\Big\{
\sum_{a,b\in\skrix}g(a,b)\varpi(a,b)-\sum_{a,b\in\skrix}e^{g(a,b)}
C(a,b)\omega(a)\omega(b)\Big\}+\sfrac{1}{2} \sum_{a,b\in\skrix}
C(a,b)\omega(a)\omega(b).
\end{aligned}
\end{equation}

Using the substitution $h=e^{g} \, \frac{C \omega \otimes
\omega}{\varpi}$ and  $\sup_{x>0} \log x - x = -1$ we obtain  the
expression
\begin{equation}\label{equ.solve}
\begin{aligned} & \sup_{g\in\skric_{2}}\Big\{
\sum_{a,b\in\skrix}g(a,b)\varpi(a,b)-\sum_{a,b\in\skrix} e^{g(a,b)}
C(a,b)\omega(a)\omega(b)\Big\}\\
&\sup_{\heap{h\in\skric_{2}}{h \ge 0}} \sum_{a,b\in\skrix} \Big[
\log \Big( h(a,b) \frac{\varpi(a,b)}{C(a,b)\omega(a) \omega(b)}\Big)
-h(a,b) \Big]  \, \varpi(a,b) \\
& = \sup_{\heap{h\in\skric_{2}}{h \ge 0}} \sum _{a,b\in\skrix} \big(
\log h(a,b) - h(a,b) \big) \, \varpi(a,b) +
\sum _{a,b\in\skrix} \log \Big( \frac{\varpi(a,b)}{C(a,b) \omega(a) \omega(b)} \Big)\, \varpi(a,b) \\
& = - \| \varpi\|  + H(\varpi \, \| \, C \omega \otimes \omega ).
\end{aligned}
\end{equation}
This gives
$\hat{I}_{1}(\omega,\,\varpi)={\mathfrak{H}_C}(\varpi\,\|\,\omega),$
which concludes the proof of the lemma.
\end{Proof}

\begin{remark} It is not hard to see that $\displaystyle{{\mathfrak
H}_C}(\cdot\,\|\,\cdot)$
is a  good rate function, as for for  all $\alpha<\infty$, its level
sets are the bounded, closed  set
$\displaystyle\{(\omega,\varpi)\in\skrim(\skrix)\times\tilde{\skrim}(\skrix\times\skrix)\colon\,{\mathfrak{H}_C}(\varpi\,\|\,\omega)\le\alpha\}$
and therefore, are compact.\end{remark}

\subsection{Lower bound in Theorem~\ref{randomge.jointL2L1L1d}(i).}
We use the LDP on the scale $n$ (but with the law $\prob$ replaced
by $\tilde{\P}$) to establish the lower bound  for some open set
$O\subset\skrim(\skrix)\times\tilde{\skrim}_{*}(\skrix\times\skrix).$

\begin{lemma}\label{randomg.lowbound1sup}

For every open set
$O\subset\skrim(\skrix)\times\tilde{\skrim}_{*}(\skrix\times\skrix).$
\begin{equation}
\liminf_{n\rightarrow\infty}\sfrac{1}{a_{n}n^{2}}\log\P\big\{(L^1,\,L^2)\in
O\big\}\ge -\inf_{(\tome,\,\tilde{\varpi})\in
O}I_{1}(\omega,\,\varpi).
\end{equation}
\end{lemma}
\begin{Proof}
Suppose  $(\omega,\,\varpi)\in O$ with $\varpi\ll
C\omega\otimes\omega.$ We define the function
$\tilde{f}_{\omega}\colon\skrix\rightarrow \R$ by
\begin{equation}\label{randomg.S1sup}
\begin{aligned}\nonumber
\tilde{f}_{\omega}(a)=\left\{\begin{array}{ll}\log\sfrac{\omega(a)}{\mu(a)},
&\mbox{if $\omega(a)> 0$,  }\\
0, & \mbox{otherwise,}
\end{array}\right.
\end{aligned}
\end{equation}
and the symmetric function
$\tilde{g}_{\varpi}\colon\skrix\times\skrix\rightarrow \R$ by
\begin{equation}\label{randomg.S2sup}
\begin{aligned}\nonumber
\tilde{g}_{\varpi}(a,b)=\left\{\begin{array}{ll}\log\sfrac{\varpi(a,b)}{C(a,b)\omega(a)\omega(b)},&\mbox{if $\varpi(a,b)>0$, }\\
0, & \mbox{otherwise.}
\end{array}\right.
\end{aligned}
\end{equation}

We recall that $$\displaystyle
\tilde{h}_{n}^{(1)}(a,b)=-\log\Big[1-p_{n}(a,b)+p_{n}(a,b)e^{\tilde{g}_{\varpi}(a,b)}\Big]^{1/a_n},\,
\mbox{for $a,b\in\skrix.$}$$  Define the symmetric function
$\tilde{\beta}_{\varpi}(a,b)$ by
$$\tilde{\beta}_{\varpi}(a,b):=\lim_{n\rightarrow\infty}\tilde{h}_{n}^{(1)}(a,b)=C(a,b)(1-e^{g_{\varpi}(a,b)}).
$$
 Choose $B_{\omega}^{1},B_{\varpi}^{2}$  open neighbourhoods of
$\omega,\varpi$ such that  $B_{\omega}^{1}\times
B_{\varpi}^{2}\subset O$ and for all $
(\tilde{\omega},\tilde{\varpi})\in B_{\omega}^{1}\times
B_{\varpi}^{2},$
$$\, \langle\varpi,\,\tilde{g}_{\varpi}\rangle+
\,
\langle\omega\otimes\omega,\,\tilde{\beta}_{\varpi}\rangle-\eps\le
\, \langle\tilde\varpi,\, \tilde{g}_{\varpi}\rangle+
\langle\tome\otimes\tome,\,\tilde{\beta}_{\varpi}\rangle.$$

We note that, the coloured random graph obtained from the function
$\tilde{g}_{\varpi}$ has colour law $\omega$ and connection
probabilities $\tilde{p}_n(a,b)\in[0,\,1]$ satisfying
$$\displaystyle a_n^{-1}\tilde{p}_n(a,b)\to \tilde
C(a,b):=\varpi(a,b)/(\omega(a)\omega(b)),\,\mbox{ as
$n\to\infty.$}$$ Write $\,\,\displaystyle
m:=0\wedge\min_{a\in\skrix}\tilde{\beta}_{\varpi}(a,a),$ and
$\displaystyle l:=0\wedge\max_{a\in\skrix}\tilde{f}(a).$  Now, using
(\ref{randomge.Itransform1sup}) we have  that
\begin{align*}
\P\Big\{(L^1,&\,L^2)\in O\Big\}\ge\tilde{\me}\Big\{\sfrac{d\P}{d\tilde{\P}}(X)\1_{\{(L^1,\,L^2)\in B_{\omega}^{1}\times B_{\varpi}^{2}\}}\Big\}\\
&=\tilde{\me}\Big\{e^{-n\langle L^1
,\,\tilde{f}_{\omega}\rangle-a_nn^2\,\langle\sfrac{1}{2}\,L^2,\,\tilde{g}_{\varpi}\rangle-a_nn^{2}\,
\langle\sfrac{1}{2}\,L^1\otimes
L^1,\,\tilde{h}_n^{(1)}\rangle+n^{2}\langle
\sfrac{1}{2}L_{\Delta}^{2},\,\tilde{h}_n^{(1)}\rangle}\times\1_{\{(L^1,\,L^2)\in B_{\omega}^{1}\times B_{\varpi}^{2}\}}\Big\}\\
&\ge
e^{-nl-a_nn^2\,\langle\varpi,\,\tilde{g}_{\varpi}\rangle/2-a_nn^{2}\,\langle\
\omega\otimes\omega,\,\tilde{\beta}\rangle/2+a_nm/4-a_nn^2\,
\eps/2}\times\tilde{\P}\Big\{(L^1,\,L^2)\in B_{\omega}^{1}\times
B_{\varpi}^{2}\Big\}.
\end{align*}

Therefore, we have
\begin{equation}\label{randomg.lowbound2sup}
\begin{aligned}
\liminf_{n\rightarrow\infty}\sfrac{1}{a_{n}n^{2}}&\log\P\big\{(L^1,\,L^2)\in
O\big\}\\
&\ge -\sfrac12\,\langle \tilde{g},\varpi\rangle-\sfrac12\,\langle
\tilde{\beta},\omega\otimes\omega\rangle-\eps
+\liminf_{n\rightarrow\infty}\sfrac{1}{a_{n}n^{2}}\log\tilde{\P}\big\{(L^1,\,L^2)\in
B_{\omega}^{1}\times B_{\varpi}^{2}\big\}.\nonumber
\end{aligned}
\end{equation}

The result follows once we prove that
\begin{equation}\label{randomgd.translowboundsup}
\liminf_{n\rightarrow\infty}\sfrac{1}{a_{n}n^{2}}\log\tilde{\P}\big\{(L^1,\,L^2)\in
B_{\omega}^{1}\times B_{\varpi}^{2}\big\}= 0.
\end{equation}

To conclude the proof, we  use the lower bound of  the LDP on the
scale $n$ (but with the law $\P$ replaced by $\tilde{\P}$), to prove
(\ref{randomgd.translowboundsup}). We notice from
Theorem~\ref{randomge.jointL2L1L1d}(ii) that, for any $\delta>0$ and
for large $n$ we have
$$\tilde{\P}\big\{(L^1,\,L^2)\in B_{\omega}^{1}\times
B_{\varpi}^{2}\big\}\ge e^{-n\alpha_2(\omega,\,\varpi)-n\delta}$$
 where $\alpha_2(\omega,\,\varpi)=\inf\big\{\tilde{I}_2(\tome,\,\tilde{\varpi}):(\tome,\,\tilde{\varpi})\in
B_{\omega}^{1}\times B_{\varpi}^{2}\big\}$ and

\begin{align*}
\tilde{I}_{2}(\tome,\,\tilde{\varpi})=\left\{
\begin{array}{ll}H(\,\tome\,\|\,\omega\,) & \mbox { if  $\tilde{\varpi}=\tilde{C}\tome\otimes\tome$, }\\
\infty & \mbox{otherwise.}
\end{array}\right.
\end{align*}
Therefore, we have
\begin{equation}\nonumber
\liminf_{n\rightarrow\infty}\sfrac{1}{a_{n}n^{2}}\log\tilde{\P}\big\{(L^1,\,L^2)\in
B_{\omega}^{1}\times
B_{\varpi}^{2}\big\}\ge\liminf_{n\to\infty}\sfrac{1}{a_nn}\{-\alpha_2(\omega,\,\varpi)-\delta\}=0,
\end{equation} since $a_nn\to\infty$ as $n\to\infty.$
This concludes the proof of the Lemma.


\end{Proof}

\subsection{Upper Bound in Theorem~\ref{randomge.jointL2L1L1s}(i).}To begin we  obtain the upper bound in a variational formulation. We
recall that $na_n \to 0$ for subcritical coloured graphs and write
$$Z_n(f):= \sfrac{1}{na_n}U_{na_nf}.$$ Notice $\displaystyle
\,Z(f):=\lim_{n\to\infty}Z_n(f)< \infty.$ Define for
  $(\omega,\varpi)\in\skrim(\skrix)\times\tilde{\skrim}_{*}(\skrix\times\skrix)$ the rate  $\hat{I}_{3}$ by
\begin{equation}
\begin{aligned}
\hat{I}_{3}(\omega,\varpi)=\sup_{\heap{f\in\skric_1}{\tilde{g}\in\skric_{2}}}\Big\{&\sum_{a\in\skrix}(f(a)-Z(f))\omega(a)\\
&+\sum_{a,b\in\skrix}\sfrac{1}{2}
\tilde{g}(a,b)\varpi(a,b)+\sum_{a,b\in\skrix}\sfrac{1}{2}(1-e^{\tilde{g}(a,b)})
C(a,b)\omega(a)\omega(b)\Big\}.\label{randomge.uppboundsuban}
\end{aligned}
\end{equation}

\begin{lemma}\label{randomge.L2L1uppboundsuban} For each closed set
$F\subset\skrim(\skrix)\times\tilde{\skrim}_{*}(\skrix\times\skrix),$\,
$$\qquad\limsup_{n\rightarrow\infty}\sfrac{1}{a_{n}n^{2}}\log\P\big\{(L^1,\,L^2)\in
F\big\}\le-\inf_{(\omega,\,\varpi)\in
F}\hat{I}_{3}(\omega,\,\varpi).$$
\end{lemma}
\begin{Proof}
Fix $f\in\skric_1.$ For any  $\tilde{g}\in\skric_2$ we define $
\tilde{\beta}\colon\skrix\times\skrix\rightarrow\R$ by
$\displaystyle\tilde{\beta}(a,b)=(1-e^{\tilde{g}(a,b)})C(a,b).$
Lemma~\ref{randomge.sequence1} gives
$\displaystyle\lim_{n\rightarrow\infty}\tilde{h}_n^{(1)}(a,b)=\tilde{\beta}(a,b),$
for all $a,b\in\skrix.$  We note that, for any $\delta>0$ and for
large $n$, we have
\begin{equation}\label{esiamp.equ3} \tilde{h}_n^{(1)}(a,b)\le
|\tilde{\beta}(a,b)|+\delta,\,\mbox{ for all $a,b\in\skrix$}
\end{equation}

Taking $\tilde{f}(a)=na_nf(a),\,$ for all $a\in\skrix,$ and using
\eqref{randomge.Itransform1sup} and \eqref{esiamp.equ3} we have
\begin{equation}\nonumber
\begin{aligned}
e^{na_n(\max_{a\in\skrix}|\tilde{\beta}(a,a)|+\delta)/2}\ge\int
e^{a_{n}n^{2}\langle
\sfrac{1}{2}L_{\Delta}^{2},\,\tilde{h}_n^{(2)}\rangle}d\tilde{\P}=\me\Big\{e^{a_nn^2\langle
L^1,\, f-Z_n(f)\rangle+a_{n}n^2\langle\sfrac{1}{2}\,
L^2,\,\tilde{g}\rangle+a_{n}n^{2}\langle\sfrac{1}{2}\,L^1\otimes
L^1,\,\tilde{h}_n^{(1)}\rangle}\Big\},
\end{aligned}
\end{equation}
for any $\delta>0$ and for  large $n.$ Therefore, we have
\begin{equation}\label{randomge.estPsuban}
\limsup_{n\rightarrow\infty}\sfrac{1}{a_{n}n^{2}}\log\me\Big\{e^{a_nn^2\langle
L^1,\, f-Z_n(f)\rangle+a_{n}n^2\langle\sfrac{1}{2}\,
L^2,\,\tilde{g}\rangle+a_{n}n^{2}\langle\sfrac{1}{2}L^1\otimes
L^1,\,\tilde{h}_n^{(1)}\rangle}\Big\}\le 0.
\end{equation}

Fix $\eps>0$ and take
$\hat{I}_{3}^{\eps}(\omega,\,\varpi)=\min\{\hat{I}_{3}(\omega,\,\varpi),{\eps}^{-1}\}-\eps.$
Suppose $(\omega,\,\varpi)\in F$ and choose
$f\in\skric_1,\,\tilde{g}\in\skric_2$ such that $$\displaystyle
\langle\,\omega,\, f-Z(f)\rangle+\sfrac{1}{2}\langle
\varpi,\tilde{g}\rangle+\sfrac{1}{2}\langle
\,\omega\otimes\omega,\,\tilde{\beta}\rangle
\ge\hat{I}_3^{\eps}(\omega,\varpi).\,$$

By finiteness of $\skrix,$ we can find open neighbourhoods
$B_{\omega}^{1},$ $B_{\varpi}^{2}$ of $\omega,\,\varpi$ such that
\begin{equation}\nonumber
\inf_{\tilde{\omega}\in B_{\omega}^{1},\,\tilde{\varpi}\in
B_{\varpi}^{2}}\big\{\langle\tome,\,f-Z(f)\rangle+  \langle\sfrac 12
\tilde{g},\,\tilde{\varpi}\rangle+\langle\sfrac
12\tome\otimes\tome,\,
\tilde{\beta}\rangle\big\}\ge\hat{I}_{3}^{\eps}(\omega,\,\varpi)-\eps.
\end{equation}

By Chebysheff's inequality  and (\ref{randomge.estPsuban}), we have
that
\begin{align}
\limsup_{n\rightarrow\infty}&\sfrac{1}{a_{n}n^{2}}\log\P\big\{(L^1,\,
L^2)
\in B_{\omega}^{1}\times B_{\varpi}^{2}\big\}\nonumber\\
&\le\limsup_{n\rightarrow\infty}\sfrac{1}{a_{n}n^{2}}\log
\me\Big\{e^{a_n n^2\langle L^1,\,
\tilde{f}-Z_n(f)\rangle+a_{n}n^2\langle\sfrac{1}{2}
L^2,\,\tilde{g}\rangle+{a_{n}}n^{2}\langle\sfrac{1}{2}L^1\otimes L^1
,\,\tilde{h}_n^{(1)}\rangle}\Big\}
-\hat{I}_{3}^{\eps}(\omega,\,\varpi)+\eps\nonumber\\
&\le-\hat{I}_{3}^{\eps}(\omega,\,\varpi)+\eps.\label{randomge.LDPballssuban}
\end{align}
By Lemma~\eqref{randomge.tightness} we choose $N(\eps)\in\N$
 (with $\alpha=\eps^{-1}$) such that
\begin{equation}\nonumber
\limsup_{n\rightarrow\infty}\sfrac{1}{a_{n}n^{2}}\log\P\Big\{|E|>a_{n}n^{2}N(\eps)
\Big\}\le-\eps^{-1}.
\end{equation}
Define for this $N,$   the set $K_{N(\eps)}$ by $$\displaystyle
K_{N(\eps)}=\big\{(\omega,\,\varpi)\in\skrim(\skrix)\times\tilde{\skrim}_{*}(\skrix\times\skrix):\|\varpi\|\le
2N(\eps)\big\}.$$

 Note $K_{N(\eps)}\cap F$ is compact and therefore may be covered by
finitely many sets $B_{\omega}^{1}\times B_{\varpi_ {r}}^{2},$
$r=1,...,m $  with $ (\omega_{r},\,\varpi_{r})\in F$ for
$r=1,...,m.$ Hence, we have
\begin{equation}\nonumber
\P\big\{ (L^1,\, L^2)\in F\big\}\le\sum_{r=1}^{m}\P\big\{(L^1,\,
L^2)\in B_{\omega_r}^{1}\times B_{\varpi_r}^{2}\big\}+\P\big\{
(L^1,\,L^2)\not\in K_{N(\eps)}\big\}.
\end{equation}
Using (\ref{randomge.LDPballssuban})  for small enough $\eps>0,$ we
obtain

\begin{equation}\nonumber
\begin{aligned}
\limsup_{n\rightarrow\infty}\sfrac{1}{a_{n}n^{2}}\log\P\big\{(L^1,\,
L^2)\in F\big\}&\le
\max_{r=1}^{m}\limsup_{n\rightarrow\infty}\sfrac{1}{a_{n}n^{2}}\log\P\big\{
(L^1,\, L^2)\in B_{\omega_r}^{1}\times
B_{\varpi_r}^{2}\big\}-\eps^{-1}\\
& \le -\hat{I}_{3}^{\eps}(\omega,\,\varpi)+\eps.
\end{aligned}
\end{equation}
Taking $\eps\downarrow 0$ we have the required statement.
\end{Proof}

We identify the rate function by solving the variational problem in
the right side of equation (\ref{randomge.uppboundsuban}).

\begin{lemma}\label{randomge.Vratesuban}
 For any
$(\omega,\,\varpi)\in\skrim(\skrix)\times\tilde{\skrim}_{*}(\skrix\times\skrix)$,
we have $\hat{I}_{3}(\omega,\,\varpi)= I_{3}(\omega,\,\varpi).$

\end{lemma}

\begin{Proof} (i)  Suppose $\omega\in\skrim(\skrix)$ is not equal $\mu.$ Define the function
$f$  by
$$
f(a)=K\log(|\omega(a)-\mu(a)|+1), \,\mbox{ for $a\in\skrix $ and
$K\in\R$.}$$

Set $\tilde{g}(a,b)=0\,$ for all $a,b\in\skrix$ in
\eqref{randomge.uppboundsuban} and note that by the choice of  $f$
we have

$$\begin{aligned}
\sum_{a\in\skrix}(f(a)-Z(f))\omega(a)
&+\sum_{a,b\in\skrix}\sfrac{1}{2}
\tilde{g}(a,b)\varpi(a,b)\sum_{a,b\in\skrix}\sfrac{1}{2}(1-e^{\tilde{g}(a,b)})C(a,b)\omega(a)\omega(b)\\
&\ge K\,
\Big(\sum_{a\in\skrix}\log(|\omega(a)-\mu(a)|+1)\omega(a)-\max_{a}|\omega(a)-\mu(a)|-1\Big)\xrightarrow{|K|\uparrow
\infty}\infty,
\end{aligned}$$

where the sign of $|K|$ is  such that last expression always stays
positive.  Suppose  $\varpi\not\ll C\omega\otimes\omega\,.$ Then
there exists $a_{0},\,b_{0}\in\skrix$ with
$C(a_{0},b_{0})\omega(a_{0})\omega(b_{0})=0$ and
$\varpi(a_{0},b_{0})>0.$ For this $(a_{0},b_{0})$ we define the
function $\tilde{g}$ by
\begin{equation}\nonumber
\tilde{g}(a,b)=\log(K(\1_{(a_{0},b_{0})}(a,b)+\1_{(b_{0},a_{0})}(a,b))+1),
\end{equation}
for $a,b\in\skrix$ and $K>0.$ Considering our $\tilde{g}$ in
\eqref{randomge.uppbounddsup} we have
\begin{equation}\nonumber
\begin{aligned}
\sum_{a,b\in\skrix}
\sfrac{1}{2}\tilde{g}(a,b)\varpi(a,b)+\sum_{a,b\in\skrix}\sfrac{1}{2}(1-e^{\tilde{g}(a,b)}C(a,b)\omega(a)\omega(b)=\log(
K+1)(\varpi(a_{0},b_{0}))\xrightarrow{K\uparrow\infty}\infty.
\end{aligned}
\end{equation}

Suppose that  $\varpi\ll C\omega\otimes\omega.$ Then, we have
$$\begin{aligned}
\hat{I}(\omega,\varpi) \ge\sfrac{1}{2} \sup_{g\in\skric_{2}}\Big\{
\sum_{a,b\in\skrix}g(a,b)\varpi(a,b)-\sum_{a,b\in\skrix} e^{g(a,b)}
C(a,b)\omega(a)\omega(b)\Big\} +\sfrac{1}{2} \sum_{a,b\in\skrix}
C(a,b)\omega(a)\omega(b).
\end{aligned}
$$

By  the substitution $h=e^{g} \, \frac{C \omega \otimes
\omega}{\varpi}$ and  $\sup_{x>0} \log x - x = -1$ we have
\eqref{equ.solve},
which yields
$$\hat{I}_{3}(\omega,\,\varpi)={I}_{3}(\omega,\,\varpi).$$

\end{Proof}

\subsection{Lower bound in Theorem~\ref{randomge.jointL2L1L1s}(i).}

\begin{lemma}\label{randomg.lowbound1sup}

For every open set
$O\subset\skrim(\skrix)\times\tilde{\skrim}_{*}(\skrix\times\skrix).$
\begin{equation}
\liminf_{n\rightarrow\infty}\sfrac{1}{a_{n}n^{2}}\log\P\Big\{(L^1,\,L^2)\in
O\Big\}\ge -\inf_{(\tome,\,\tilde{\varpi})\in
O}I_{3}(\omega,\,\varpi).
\end{equation}
\end{lemma}
\begin{Proof}
Suppose  $(\omega,\,\varpi)\in O$ with $\varpi\ll
C\omega\otimes\omega$ and $\omega=\mu.$ Take $\tilde{f}(a)=0,\,$
\,for all $a\in\skrix.$ Define the symmetric function
$\tilde{g}_{\varpi}\colon\skrix\times\skrix\rightarrow \R$ by
\begin{equation}\label{randomg.S2sup}
\begin{aligned}\nonumber
\tilde{g}_{\varpi}(a,b)=\left\{\begin{array}{ll}\log\sfrac{\varpi(a,b)}{C(a,b)\omega(a)\omega(b)},&\mbox{if $\varpi(a,b)>0$, }\\
0, & \mbox{otherwise.}
\end{array}\right.
\end{aligned}
\end{equation}

Recall that $$\displaystyle
\tilde{h}_{n}^{(1)}(a,b)=-\log\Big[1-p_{n}(a,b)+p_{n}(a,b)e^{\tilde{g}_{\varpi}(a,b)}\Big]^{1/a_n},\,
\mbox{for $a,b\in\skrix.$}$$

 Define the function
$\tilde{\beta}_{\varpi}(a,b)$ by
$$\displaystyle\tilde{\beta}_{\varpi}(a,b):=\lim_{n\rightarrow\infty}\tilde{h}_{n}^{(1)}(a,b)=C(a,b)(1-e^{g_{\varpi}(a,b)}).
$$ Choose $B_{\omega}^{1},B_{\varpi}^{2}$  open neighbourhoods of
$\omega,\varpi$ such that  $B_{\omega}^{1}\times
B_{\varpi}^{2}\subset O$ and for all
$(\tilde{\omega},\tilde{\varpi})\in B_{\omega}^{1}\times
B_{\varpi}^{2},$
$$\, \langle\varpi,\,\tilde{g}_{\varpi}\rangle+
\,
\langle\omega\otimes\omega,\,\tilde{\beta}_{\varpi}\rangle-\eps\le
\, \langle\tilde\varpi,\, \tilde{g}_{\varpi}\rangle\,+
\langle\tilde{\omega}\otimes\tilde{\omega},\,\tilde{\beta}_{\varpi}\rangle.$$

We note that, the random  coloured graph obtained from the function
$\tilde{g}_{\varpi}$ has colour law $\omega$ and connection
probabilities satisfying $$\displaystyle a_n^{-1}\tilde{p}_n(a,b)\to
\tilde{C}(a,b):=\varpi(a,b)/(\omega(a)\omega(b)),\,\mbox{ as
$n\to\infty.$}$$

Using  (\ref{randomge.Itransform3super}) we have

\begin{align*}
\P\Big\{(L^1,\,L^2)\in O\Big\}&\ge\tilde{\me}\Big\{\sfrac{d\P}{d\tilde{\P}}(X)\1_{\{(L^1,\,L^2)\in B_{\omega}^{1}\times B_{\varpi}^{2}\}}\Big\}\\
&=\tilde{\me}\Big\{e^{-a_nn^2\,\langle\sfrac{1}{2}\,L^2,\,\tilde{g}_{\varpi}\rangle-a_nn^{2}\,
\langle\sfrac{1}{2}\,L^1\otimes L^1,\,\tilde{h}_n^{(1)}\rangle+a_n
n^{2}\,\langle
L_{\Delta}^{2},\,\tilde{h}_n^{(1)}\rangle}\times\1_{\{(L^1,\,L^2)\in B_{\omega}^{1}\times B_{\varpi}^{2}\}}\Big\}\\
&\ge
e^{-a_nn^2\langle\sfrac12\,\varpi,\,\tilde{g}_{\varpi}\rangle-a_nn^{2}\langle\sfrac12\,\omega\otimes\omega,\,\tilde{\beta}\rangle+a_nm/4-\sfrac12
\,a_nn^2 \eps}\times\tilde{\P}\Big\{(L^1,\,L^2)\in
B_{\omega}^{1}\times B_{\varpi}^{2}\Big\},
\end{align*}
where $m:=0\wedge\min_{a\in\skrix}\tilde{\beta}_{\varpi}(a,a).$
Therefore, we have
\begin{equation}\label{randomg.lowbound2suban}
\begin{aligned}
\liminf_{n\rightarrow\infty}\sfrac{1}{a_{n}n^{2}}\log\P\big\{&(L^1,\,L^2)\in
O\big\}\\
&\ge -\sfrac12\,\langle\,\varpi,\,
\tilde{g}\rangle-\sfrac12\,\langle \omega\otimes\omega,\,
\tilde{\beta}\rangle-\eps
+\liminf_{n\rightarrow\infty}\sfrac{1}{a_{n}n^{2}}\log\tilde{\P}\big\{(L^1,\,L^2)\in
B_{\omega}^{1}\times B_{\varpi}^{2}\big\}.\nonumber
\end{aligned}
\end{equation}

The result follows once we prove that
\begin{equation}\label{randomgd.translowboundsuban}
\liminf_{n\rightarrow\infty}\sfrac{1}{a_{n}n^{2}}\log\tilde{\P}\big\{(L^1,\,L^2)\in
B_{\omega}^{1}\times B_{\varpi}^{2}\big\}= 0.
\end{equation}

We  use the upper bound (but now with the law  $\P$ replaced by
$\tilde{\P}$)  to prove
(\ref{randomgd.translowboundsuban}).\\

Therefore, we have
\begin{equation}\nonumber
\limsup_{n\rightarrow\infty}\sfrac{1}{a_nn^2}\log\tilde{\P}\big\{(L^1,\,L^2)\in
(B_{\omega}^{1}\times B_{\varpi}^{2})^{c}\big\}\\
\le-\inf_{(\tome,\,\tilde{\varpi})\in
\hat{F}}\tilde{I}_3(\tome,\,\tilde{\varpi}),
\end{equation}

\begin{align*}
\tilde{I}_{3}(\tome,\,\tilde{\varpi})=\left\{
  \begin{array}{ll}\sfrac{1}{2}{\mathfrak{H}_{\tilde C}}(\tilde{\varpi}\,\|\,\omega) & \mbox { if $\tome=\omega$, }\\
\infty & \mbox{otherwise.}
\end{array}\right.
\end{align*}

where   $\hat{F}=(B_{\omega}^{1}\times B_{\varpi}^{2})^c$ and
$(B_{\omega}^{1}\times B_{\varpi}^{2})^c$ is the complement of the
set $B_{\omega}^{1}\times B_{\varpi}^{2}.$  It remain  for us to
show that the infimum is positive. Suppose by contradiction there
exists the sequence $(\omega_n,\,\varpi_n)\in \hat{F}$ such that
$\tilde{I}_{3}(\tome,\,\tilde{\varpi})\downarrow 0.$ Then, because
$\tilde{I}_{3}$  is good rate function with all its level sets
compact, and  by lower semicontinuity  of the mapping
$(\tome,\,\tilde{\varpi})\to \tilde{I}_{3}(\tome,\,\tilde{\varpi})$,
we can construct a limit point $(\tilde{\varpi},\,\tome)\in\hat{F}$
with $\tilde{I}_{3}(\tome,\,\tilde{\varpi})=0.$ This means
$\tome=\omega$ and
$\tilde{\varpi}=\tilde{C}\tome\otimes\tome=\varpi,$ and hence,
contradicting $(\tome,\,\tilde{\varpi})\in\hat{F}.$
\end{Proof}

\subsection{Upper bound in Theorems~\ref{randomge.jointL2L1L1s}(ii).}
We define for
$(\omega,\varpi)\in\skrim(\skrix)\times\tilde{\skrim}_{*}(\skrix\times\skrix),$
  the function  $\hat{I}_{4}(\omega,\,\varpi)$ by
\begin{equation}
\hat{I}_{4}(\omega,\,\varpi)=\sup_{\tilde{f}\in\skric_1}\Big\{\sum_{a\in\skrix}\big(\tilde{f}(a)-U_{\tilde{f}}\big)\omega(a)\Big\}\\
\label{randomge.uppboundnsubn}
\end{equation}

\begin{lemma}\label{randomge.L2L1uppboundnsuban}
 For each closed set
$F\subset\tilde{\skrim}_{*}(\skrix\times\skrix),$ we have
\begin{equation}
\limsup_{n\rightarrow\infty}\sfrac{1}{n}\log\P\big\{(L^1,\,L^2)\in
F\big\}\le-\inf_{(\omega,\,\varpi)\in
F}\hat{I}_{4}(\omega,\,\varpi).\nonumber
\end{equation}
\end{lemma}
\begin{Proof}
Fix $\tilde{f}\in\skric_1$ and take  $\tilde{g}(a,b)=0,$ for all
$a,b\in\skrix.$ We observe that by this choice of  $\tilde{g}$
$$\displaystyle
\tilde{h}_n^{(2)}(a,b)=0,\,\mbox{for all $a,b\in\skrix.$}$$

Using (\ref{randomge.Itransform3super}) we obtain
$\me\big\{e^{n\langle L^1,\,
\tilde{f}-U_{\tilde{f}}\rangle}\big\}=\int \,d\tilde{\prob}\le 1 \,$
and  therefore, we have
\begin{equation}\label{randomg.estPnsubn}
\limsup_{n\rightarrow\infty}\sfrac{1}{n}\log\me\Big\{e^{n\langle
\,L^1,\tilde{f}-U_{\tilde{f}}\rangle}\Big\}\le 0.
\end{equation}

Now fix $\eps>0$ and write
$\hat{I}_{4}^{\eps}(\omega,\,\varpi):=\min\{\hat{I}_{4}(\omega,\,\varpi),{\eps}^{-1}\}-\eps.$
We suppose $(\omega,\,\varpi)\in F$ and choose
$\tilde{f}\in\skric_1$ such that
$$\langle\tilde{f}-U_{\tilde{f}},\omega\rangle\ge\hat{I}_{4}^{\eps}(\omega,\,\varpi).$$
By finiteness of $\skrix$,  we can find  open neighbourhoods
$B_{\varpi}^{2}$ and $B_{\omega}^{1}$ of $\varpi,\,\omega$ such that
\begin{equation}\nonumber
\inf_{\tilde{\omega}\in B_{\omega}^{1}}\big\{\langle\tome,\,
\tilde{f}-U_{\tilde{f}} \rangle\big\}\ge
\hat{I}_{4}^{\eps}(\omega,\varpi)-\eps.
\end{equation}
Using  Chebysheff's inequality  and (\ref{randomg.estPnsubn}), we
have that

\begin{equation}
\begin{aligned}
\limsup_{n\rightarrow\infty}\sfrac{1}{n} \log\prob\big\{(L^1,\,
L^2)\in B_{\omega}^{1}\times
B_{\varpi}^{2}\big\}&\le\limsup_{n\rightarrow\infty}\sfrac{1}{n}\log
\me\Big\{e^{n\langle L^1,\,\tilde{f}-U_{\tilde{f}}\rangle}\Big\}
-\hat{I}_{4}^{\eps}(\omega,\varpi)+\eps\\
&\le-\hat{I}_{4}^{\eps}(\omega,\varpi)+\eps.\label{randomg.LDPballssuban}
\end{aligned}\end{equation}

 By Lemma~\ref{randomge.tightness} with
 $\alpha=\eps^{-1},$ we choose $N(\eps)\in\N$  such that
\begin{equation}\nonumber
\limsup_{n\rightarrow\infty}\sfrac{1}{n}\log\P \Big\{|E|>a_{n}n^{2}
N(\eps) \Big\}\le-\eps^{-1}.
\end{equation}

 We define for this $N$  the set $K_{N(\eps)}$ by
$$\displaystyle
K_{N(\eps)}=\big\{(\omega,\,\varpi)\in\skrim(\skrix)\times\tilde{\skrim}_{*}(\skrix\times\skrix):\|\varpi\|\le
2N(\eps)\big\}.$$

The set $K_{N(\eps)}\cap F$ is compact and therefore may be covered
by finitely many sets

$B_{\omega_ {r}}^{1}\times B_{\varpi_ {r}}^{2},$ $r=1,\ldots ,m$
with $ (\omega_{r},\varpi_{r})\in F$ for $r=1,\ldots ,m.$  Hence, we
have
\begin{equation}\nonumber
\prob\big\{(L^1,\, L^2)\in F\big\}\le\sum_{r=1}^{m}\prob\big\{(L^1,
L^2)\in B_{\omega_{r}}^{1}\times
B_{\varpi_{r}}^{2}\big\}+\prob\big\{(L^1, L^2)\not\in
K_{N(\eps)}\big\}.
\end{equation}

Now we use (\ref{randomg.LDPballssuban})  to obtain, for all
sufficiently small $\eps>0$,
\begin{equation}\nonumber
\begin{aligned}
\limsup_{n\rightarrow\infty}\sfrac{1}{n}\log\prob\big\{(L^1, L^2)\in
F\big\}&\le
\max_{r=1}^{m}\limsup_{n\rightarrow\infty}\sfrac{1}{n}\log\prob\big\{(L^1,
L^2)\in B_{\omega_{r}}^{1}\times B_{\varpi_{r}}^{2}\big\}-\eps^{-1}\\
&  \le -\inf_{(\omega,\varpi)\in
F}\hat{I}_{4}^{\eps}(\omega,\varpi)+\eps.
\end{aligned}
\end{equation}
Taking $\eps\downarrow 0$ we have the desired statement.
\end{Proof}

By the variational characterization of the relative entropy we have
$I_{4}(\omega,\, \varpi)=H(\omega\,\|\,\mu),$ which ends the proof
of the upper bound.

\subsection{Lower bound in Theorems~\ref{randomge.jointL2L1L1s}(ii)}

\begin{lemma}\label{randomg.losnsubn}
For every open set $
O\subset\skrim(\skrix)\times\tilde{\skrim}_{*}(\skrix\times\skrix),$
\begin{equation}
\liminf_{n\rightarrow\infty}\sfrac{1}{n}\log\P\Big\{(L^1,L^2)\in
O\Big\}\ge -\inf_{(\omega,\,\varpi)\in O} I_{4}(\omega,\,\varpi).
\end{equation}
\end{lemma}
\begin{Proof}
Suppose  $(\omega,\,\varpi)\in O$ with $\varpi\ll
C\omega\otimes\omega.$ We define the function
$\tilde{f}_{\omega}\colon\skrix\rightarrow \R$ by
\begin{equation}\label{randomg.S1supbn}
\begin{aligned}\nonumber
\tilde{f}_{\omega}(a)=\left\{\begin{array}{ll}\log\sfrac{\omega(a)}{\mu(a)},
&\mbox{if $\omega(a)> 0$,}\\
0, & \mbox{otherwise,}
\end{array}\right.
\end{aligned}
\end{equation}
and the symmetric function
$g_{\varpi}\colon\skrix\times\skrix\rightarrow \R$ by
\begin{equation}\label{randomg.S2sup}
\begin{aligned}\nonumber
g_{\varpi}(a,b)=\left\{\begin{array}{ll}\log\sfrac{\varpi(a,b)}{C(a,b)\omega(a)\omega(b)},&\mbox{if $\varpi(a,b)>0$, }\\
0, & \mbox{otherwise.}
\end{array}\right.
\end{aligned}
\end{equation}

Set $\tilde{g}_{\tilde\varpi}(a,b)=na_ng_{\varpi}(a,b),\,$ for all
$a,b\in\skrix$ and  note by  the choice of $\tilde{g}$  we have
$$\lim_{n\rightarrow\infty}\tilde{h}_{n}^{(2)}(a,b)=0,\,\mbox{ for all $a,b\in\skrix.$}$$
Choose $B_{\omega}^{1},B_{\varpi}^{2}$ open neighbourhoods of
$\omega,\,\varpi$ such that $B_{\omega}^{1}\times
B_{\varpi}^{2}\subset O$ and for all
$(\tilde{\omega},\tilde{\varpi})\in B_{\omega}^{1}\times
B_{\varpi}^{2},$ $$\langle
\,\omega,\,\tilde{f}_{\omega}\rangle+\sfrac12\,\|\varpi\|\eps-\eps\le
\langle
\tome,\,\tilde{f}_{\omega}\rangle+\sfrac12\,\|\tilde{\varpi}\|\eps.$$

We use the  probability measure $\tilde{\P}$ given by
$\tilde{g}_{\varpi}.$  We observe that the colour law is $\omega$
and the connection probabilities satisfy
$a_n^{-1}\tilde{p}_n(a,b)\to
\tilde{C}(a,b):=\varpi(a,b)/(\omega(a)\omega(b)),$ as $n$ approaches
infinity. Therefore, using (\ref{randomge.Itransform3super}) we have
that
\begin{equation}\nonumber
\begin{aligned}
\prob\Big\{(L^1,L^2) \in
O\Big\}&\ge\tilde{\me}\Big\{\sfrac{d\prob}{d\tilde{\prob}}(X)\one_{\{(L^1,L^2)\in
B_{\omega}^{1}\times B_{\varpi}^{2}\}}\Big\}\\
&=\tilde{\me}\Big\{ e^{-n\langle
L^1,\tilde{f}_{\omega}\rangle-n\langle \,\sfrac12
L^2,\,\tilde{g}\rangle-n\langle \,\sfrac12 L^1\otimes
L^1,\,\tilde{g}\rangle+n\langle \,
L_{\Delta}^1,\,\tilde{h}_n^{(2)}\rangle}\one_{\{(L^1,\,L^2)\in
B_{\omega}^{1}\times B_{\varpi}^{2}\}}\Big\}\\
 &\ge
e^{-n\,\langle
\omega,\,\tilde{f}_{\omega}\rangle-n\,\|\varpi\|\eps/2-n\,\eps-\,a_nn^2l/2+o(1)}\times\tilde{\prob}\Big\{(L^1,L^2)\in
B_{\omega}^{1}\times B_{\varpi}^{2}\Big\},
\end{aligned}
\end{equation}
where $\displaystyle l=0\wedge
\max_{a,b\in\skrix}\tilde{g}_{\tilde\varpi}(a,b).$ Therefore, we
have that
\begin{equation}\label{randomg.lowbound2subn}
\begin{aligned}
\liminf_{n\rightarrow\infty}\sfrac{1}{n}\log\prob\big\{(L^1,L^2)\in
O\big\} \ge-\langle \omega,\,
\tilde{f}_{\omega}\rangle-\sfrac12\,\|\varpi\|\eps-\eps
+\liminf_{n\rightarrow\infty}\sfrac{1}{n}\log\tilde{\prob}
\big\{(L^1,L^2)\in B_{\omega}^{1}\times
B_{\varpi}^{2}\big\}.\nonumber
\end{aligned}
\end{equation}

The result follows once we prove that
\begin{equation}\label{randomg.translowboundnsubn}
\liminf_{n\rightarrow\infty}\sfrac{1}{n}\log\tilde{\prob}\big\{(L^1,L^2)\in
B_{\omega}^{1}\times B_{\varpi}^{2}\big\}= 0.
\end{equation}

We use the lower bound  of the LDP on the scale $a_nn^2$ (but now
with the law $\P$ replaced by $\tilde{\P}$) to prove
(\ref{randomg.translowboundnsubn}). We observe from the lower bound
of Theorem~\ref{randomge.jointL2L1L1s}(i) that, for any $\delta>0$
and for large $n$ we have
$$\tilde{\P}\big\{(L^1,\,L^2)\in B_{\omega}^{1}\times
B_{\varpi}^{2}\big\}\ge
e^{-a_nn^2(\alpha_3(\omega,\,\varpi)+\delta)},$$
 where $\alpha_3(\omega,\,\varpi)=\inf\big\{\tilde{I}_3(\tome,\,\tilde{\varpi}):(\tome,\,\tilde{\varpi})\in
B_{\omega}^{1}\times B_{\varpi}^{2}\big\}$ and
\begin{align*}
\tilde{I}_{3}(\tome,\,\tilde{\varpi})=\left\{
  \begin{array}{ll}\sfrac{1}{2}{\mathfrak{H}_{\tilde C}}(\tilde{\varpi}\,\|\,\omega) & \mbox { if $\tome=\omega$, }\\
\infty & \mbox{otherwise.}
\end{array}\right.
\end{align*}

Therefore, we have that
\begin{equation}\nonumber
\liminf_{n\rightarrow\infty}\sfrac{1}{n}\log\tilde{\P}\big\{(L^1,\,L^2)\in
B_{\omega}^{1}\times
B_{\varpi}^{2}\big\}\ge\liminf_{n\to\infty}a_{n}{n}\{-\alpha_3(\omega,\,\varpi)-\delta\}=0,
\end{equation}
since $a_nn\to 0$  as $n$ approaches infinity. This  concludes the
proof of the Lemma.

\end{Proof}

\subsection{Proof of Theorem~\ref{smb.tree}.}Recall $N_0$ from the boundedness of ${\mathbb{Q}} $ and
write
 $$\skrix_{0}^{*}=\bigcup_{n=0}^{N_0}\{n\}\times\skrix^{n}.$$
We equip it with the discrete topology. We recall  also that $\pi$
is the unique eigenvector (normailzed to a probability vector) of
the matrix $A$ corresponding to the eigenvalue $1.$ We recall that
$\prob_n$ is the law of a multitype Galton-Watson tree conditioned
to have $n$ vertices, 
and  derive from Theorem~\ref{multgaltonwatsontree.DMS03} the
following weak law of large numbers.\\

\begin{lemma}\label{weaklaw.tree}
Suppose that $X$ is an irreducible, critical multitype Galton-Watson
tree with  bounded offspring law ${\mathbb{Q}},$  conditioned  to
have exactly $n$ vertices. Then, for any $\eps>0$
\begin{equation}\label{weaklaw.eq}
\lim_{n\to\infty}
\prob_n\Big\{\max_{(a,c)\in\skrix\times\skrix_{0}^{*}}\big|M_X(a,c)-\pi\otimes{\mathbb{Q}}(a,c)\big|\ge
\eps\Big\}=0.
\end{equation}
\end{lemma}

\begin{Proof}
Define the closed set
$$F=\big\{\,\nu\in\skrim(\skrix\times\skrix_0) :  \max_{(a,c)\in\skrix\times\skrix_{0}^{*}}\big|\nu(a,c)-\pi\otimes{\mathbb{Q}}(a,c)\big|\ge
\eps\,\big\}.$$

We observe from Theorem~\ref{multgaltonwatsontree.DMS03} that ,
\begin{equation}\label{smb.LDU}
\limsup_{n\rightarrow\infty}\sfrac1n \log\prob_n\big\{\,M_X\in
F\,\}\le-\inf_{\nu\in F}J(\nu).
\end{equation}

We show by contradiction that the right hand side of \eqref{smb.LDU}
is negative.\\

 To do this, we suppose that there exists sequence
$\nu_n$ in $F$ such that $J(\nu_n)\downarrow 0.$ Then, because $J$
is a good rate function and its level sets are compact, and  by
lower semicontinuity of the mapping $\nu\mapsto J(\nu),$ there is a
limit $\nu\in F$ with $J(\nu)=0.$ Hence, we have that $\nu$ is
shift-invariant and $H(\nu\,\|\,\nu_1\otimes{\mathbb{Q}})=0.$ This
implies $\nu(a,c)=\nu_1\otimes{\mathbb{Q}}(a,c),$ for every
$(a,c)\in\skrix\times\skrix_{0}^{*}.$   Using shift-invariance of
$\nu,$ for any $b\in\skrix,$ we have
$$\sum_{a\in\skrix}A(b,a)\nu_1(a)=\sum_{(a,c)\in\skrix\times\skrix_0^{*}}{\mathbb{Q}}\{c\,|\,a\}m(b,c)\nu_1(a)=\sum_{(a,c)\in\skrix\times\skrix_0^{*}}\nu(a,c)m(b,c)=\nu_1(b).$$
This means that $\nu_1$ is a nonnegative eigenvector of $A.$
 By uniqueness of the Perron-Frobenius eigenvector, see, for example
Dembo \emph{et al.}  \cite[Theorem~3.1.1(d)]{DZ98}, we infer that
$\nu_1=\pi.\,$ This contradicts $\pi\otimes{\mathbb{Q}}\not\in F.$
\end{Proof}

We recall that $\skrit$ is set of all finite rooted planar trees
$T$,  $V=V(T)$ is set of all vertices and $|T|$ is the number of
vertices in the tree $T.$ We now compute the probability weight
$P_n(x)$  of $x\in\skrit$ as

\begin{align*}
P_n(x) =\sfrac{\mu(x(\rho))}{\prob\{|T|=n\}} \prod_{v\in
V(T),|T|=n}{\mathbb{Q}}\big\{C(v)=c(v)\,|\,X(v)=x(v)\big\},
\end{align*}
$(x(v),c(v))$ is the type, and the configuration of  children of
vertex $v$ of $x\in\skrit.$  Therefore, we have that
$$-\sfrac1n \log
P_n(x)=-\sfrac1n\log\mu(x(\rho))+\sfrac1n\log\prob\{|T|=n\}+\langle
M_x,\,-\log{\mathbb{Q}}\rangle.$$

\pagebreak

Now the  term $\sfrac1n\log\mu(x(\rho))$ converges to zero, while
 the term $\sfrac1n\log\prob\{|T|=n\}$  converges to zero because ${\mathbb{Q}}$ is bounded. See Dembo et al. \cite[Lemma~3.1]{DMS03}.
We observe that  $-\log{\mathbb{Q}}$ is almost surely bounded on the
support of $M_X$ and therefore, by
 Lemma~\ref{weaklaw.tree} we have
 $\displaystyle\langle
M_x,\,-\log{\mathbb{Q}}\rangle\rightarrow \langle
\pi\otimes{\mathbb{Q}} ,\,-\log{\mathbb{Q}}\rangle,$  which
concludes the proof of Theorem~\ref{smb.tree}.

\subsection{Derivation of Theorems~\ref{randomgsmb.sparse}~and~\ref{randomgsmb.criticals}.}\label{DMT}\label{PAEPSHS}

We recall that $\prob_n$ is the law of 
a coloured random graph with $n$ vertices, and derive from our large
deviation principles for coloured random graphs the following weak
law of large numbers.

\begin{lemma}\label{WLLN}
Suppose that $X$ is a  coloured random graph with colour law
 $\mu\colon\skrix\rightarrow (0,1]$ and
 connection probabilities $p_{n}:\skrix\times\skrix\rightarrow[0,1]$ such
 that $a_n^{-1}p_n(a,b) \to C(a,b)$ for some  sequence $(a_n)$ with
 $a_nn\rightarrow 0$  or
 $a_nn\rightarrow1$ or $a_nn\rightarrow\infty$
 and $C:\skrix\times\skrix\rightarrow[0,\infty)$ nonzero. Then, for
 any $\eps>0$ we have

$$\lim_{n\to\infty} \prob_n\big\{ \sup_{a\in\skrix} |L^1(a) - \mu(a)|
 \ge \eps \big\}=0 $$  and

 $$\lim_{n\to\infty} \prob_n\big\{ \sup_{a,b\in\skrix} |L^2(a,b) - \mu(a) C(a,b) \mu(b) |\ge \eps \big\}=0.$$
 \end{lemma}

From
Theorem~\ref{main},~Theorem~\ref{randomge.jointL2L1L1d}~and~Theorem~\ref{randomge.jointL2L1L1s}
we prove this lemma.

 To begin, we define a closed set $$F_1=\big\{
(\omega,\varpi)\in\skrim(\skrix) \times
\tilde{\skrim}_{*}(\skrix\times\skrix)\colon \sup_{a,b\in\skrix}
|\varpi(a,b) - \mu(a) C(a,b) \mu(b) | \ge \eps\}.$$ We observe that
in the sparse case (when $na_{n}\to 1$), by Theorem~\ref{main},
\begin{equation}\label{randomsmb.lastb1}
\limsup_{n\rightarrow\infty}\sfrac{1}{n}\log\prob_n\Big\{(L^1,L^2)
\in F_1 \Big\}\le -\inf_{(\omega,\varpi)\in F_1} I(\omega,\varpi).
\end{equation}

We show by contradiction that the right handside of
\eqref{randomsmb.lastb1} is negative. For this purpose suppose that
there exists sequence  $(\omega_n,\varpi_n)$ in $F_1$ such that
$I(\omega_n,\varpi_n)\downarrow 0.$ Then, because $I$ is a good rate
function and its level sets are compact, and by  lower
semicontinuity of the mapping $(\omega,\varpi)\mapsto
I(\omega,\varpi)$, there is a limit point $(\omega,\varpi)\in F_1$
with $I(\omega,\varpi)=0$. By Doku  et al. \cite[Lemma~3.4]{DM06a},
we have $H(\omega\,\|\,\mu)=0$ and
${\mathfrak{H}_C}(\varpi\,\|\,\omega)=0.$ This implies
$\omega(a)=\mu(a),$ and $\varpi(a,b)=C(a,b)\omega(a)\omega(b),$ for
$a,b\in\skrix$ which contradicts $(\omega,\varpi)\in F_1$. Hence
 as required.\\

For the subcritical case we can argue similarly with the LDP on the
scale $a_n n^2$ with rate function $I_3$.  The first statement of
Lemma~\ref{WLLN}  follows similarly using the set $$F_2=\big\{
(\omega,\varpi)\in\skrim(\skrix) \times
\tilde{\skrim}_{*}(\skrix\times\skrix)\colon \sup_{a\in\skrix}
|\omega(a) - \mu(a) | \ge \eps\}$$ and the LDP of Theorem~\ref{main}
in the sparse case, and the LDP on the scale $n$ with rate function
$I_4$ in the subcritical case. Finally, in the
 supercritical case,  an analogous argument can be
carried out using $F=F_1\cup F_2$ and the LDP on the scale $n$ with
rate function $I_2$.\\

We recall that  $V$ is  a fixed set of $n$ vertices, say
$V=\{1,\ldots,n\},$  $\skrig_n$ is the set of all (simple) graphs
with vertex set $V=\{1,\ldots,n\}$ and  $\displaystyle
E\subset\skrie:=\big\{(u,v)\in V\times V \, : \, u<v\big\}$  the
edge set.

 We now compute the probability weight $P_n(x)$ of
$x\in\skrig_n$ as
\begin{align*}
P_n(x)& 
=\prod_{u\in V}\mu(x(u))\prod_{(u,v)\in
E}p_{n}(x(u),x(v))\prod_{(u,v)\not\in
E} \big( 1-p_{n}(x(u),x(v)) \big)\\
&=\prod_{u\in V}\mu(x(u))\prod_{(u,v)\in E}\sfrac
{p_{n}(x(u),x(v))}{1-p_{n}(x(u),x(v))}\prod_{(u,v)\in \skrie}
\big(1-p_{n}(x(u),x(v)) \big).
\end{align*}
Therefore, we have in the case of Theorem~\ref{randomgsmb.sparse}
\begin{align*}
 -\sfrac{1}{ a_n n^{2} \log n}\log P_n(x)=\langle
 L^1,  -\sfrac{\log\mu}{a_n \, n \,\log n}\rangle &+\sfrac 12\,
\langle L^2,-\sfrac{\log(p_{n}/(1-p_{n}))}{\log n}\rangle\\
&+\sfrac 12\,\langle L^1\otimes L^1,-\sfrac{\log(1-p_{n})}{ a_n \log
n}\rangle+\sfrac 12\, \langle
L_{\Delta}^{1},-\sfrac{\log(1-p_{n})}{a_n \, n\, \log n}\rangle.
\end{align*}
In the case of Theorem~\ref{randomgsmb.criticals} we have
\begin{align*}
 -\sfrac{1}{n}\log P_n(x)=\langle L^1,-\log\mu\rangle &+\sfrac 12\,\langle L^2,-\sfrac{\log(p_{n}/(1-p_{n}))}{\log n}\rangle\\
&+\sfrac 12\,\langle L^1\otimes L^1,-n \log(1-p_{n})\rangle+ \sfrac
12\,\langle L_{\Delta}^{1},-\log(1-p_{n})\rangle.
\end{align*}

Now in the first case the integrands $\sfrac{-\log\mu}{a_n \, n
\,\log n},\, \,\sfrac{-\log(1-p_{n})}{ a_n \log n}\,\mbox{and}\,
\sfrac{-\log(1-p_{n})}{a_n \, n\, \log n}$ all converge  to zero,
while $\sfrac{-\log(p_{n}/(1-p_{n}))}{\log n}
 \to 1.$  Hence Theorem~\ref{randomgsmb.sparse} follows from
Theorem~\ref{WLLN}.

  In the second case  both integrand $-\log(1-p_{n})\,\mbox{ and
  } -n\, \log(1-p_{n})$
converges to zero.  Therefore, Theorem~\ref{randomgsmb.criticals}
follows from Lemma~\ref{WLLN}.

\bigskip
{\bf Acknowledgments:} This paper contains material from my PhD
thesis (Bath). I would like to thank LMS  for supporting my recent
visit to Bath.



\end{document}